\documentclass[twocolumn,aps,epsfig,nofootinbib]{revtex4}

\usepackage{graphicx}
\usepackage{epstopdf}
\usepackage{latexsym}
\usepackage{amssymb}
\usepackage{amsmath}
\usepackage{color}
\usepackage{mathrsfs}
\usepackage[center]{subfigure}

\begin{document}

\newcommand{\bq}{\begin{equation}}
\newcommand{\eq}{\end{equation}}
\newcommand{\bqn}{\begin{eqnarray}}
\newcommand{\eqn}{\end{eqnarray}}
\newcommand{\nb}{\nonumber}
\newcommand{\lb}{\label}
\newcommand{\PRL}{Phys. Rev. Lett.}
\newcommand{\PL}{Phys. Lett.}
\newcommand{\PR}{Phys. Rev.}
\newcommand{\PRD}{Phys. Rev. D}
\newcommand{\CQG}{Class. Quantum Grav.}
\newcommand{\JCAP}{J. Cosmol. Astropart. Phys.}
\newcommand{\JHEP}{J. High. Energy. Phys.}
\newcommand{\PLB}{Phys. Lett. B}

\title{Power spectra and spectral indices of $k$-inflation: High-order corrections}

\author{Tao Zhu, Anzhong Wang}
\affiliation{Institute for Advanced Physics $\&$ Mathematics, Zhejiang University of Technology, Hangzhou, 310032, China\\
GCAP-CASPER, Physics Department, Baylor University, Waco, TX 76798-7316, USA}

\author{Gerald Cleaver}
\affiliation{EUCOS-CASPER, Physics Department, Baylor University, Waco, TX 76798-7316, USA}

\author{Klaus Kirsten and Qin Sheng}
\affiliation{GCAP-CASPER, Mathematics Department, Baylor University, Waco, TX 76798-7328, USA}

\date{\today}

\begin{abstract}
$k$-inflation represents the most general single-field inflation, in which the perturbations usually obey an equation of motion with a time-dependent sound speed. In this paper, we study the observational predictions of the $k$-inflation
by using the high-order uniform asymptotic approximation method. We calculate explicitly the slow-roll expressions of the power spectra, spectral indices, and running of the spectral indices for both the scalar and tensor perturbations.
These expressions are all written in terms of the Hubble and sound speed flow parameters. It is shown that the previous results obtained by using the first-order uniform asymptotic approximation have been significantly improved by the high-order corrections of the uniform asymptotic approximations. Furthermore, we also check our results by comparing them with the ones obtained by other approximation methods, including  the Green's function method, WKB approximation, and improved WKB approximation, and find the relative errors.
\end{abstract}

\maketitle

\section{Introduction}
\renewcommand{\theequation}{1.\arabic{equation}} \setcounter{equation}{0}

Inflationary cosmology has become the dominant paradigm for describing the evolution of the very early universe. It not only solves several fundamental and conceptual problems of the conventional big bang cosmology, but also provides an elegant mechanism for generating the primordial density perturbations and primordial gravitational waves (PGWs) \cite{Guth,InfGR}. Both the density perturbations and PGWs create the cosmic microwave background temperature anisotropies, which have already been detected by various CMB observations, such as WMAP \cite{WMAP}, Planck \cite{PLANCK}, and BICEP2 \cite{BICEP2}. The observational results from WMAP, Planck, BICEP2, and also forthcoming CMB experiments provide measurements of the primordial power spectra and spectral indices more accurate than ever before. For this, the comparison of inflationary models with accurate  observations requires precise  theoretical predictions.

However, in general it is impossible to obtain exact spectra and spectral indices analytically, and thus one has to use some approximate methods. In particular, using the slow-roll approximation, the spectra of both scalar and tensor perturbations were first calculated to the first-order in the slow-roll approximations in \cite{Stewart1993PLB}, in which  the slow-roll parameters were assumed to be small and constant. Beyond the first order slow-roll approximation, one expects the time-dependence of the slow-roll parameters could contribute to the spectra at the second-order in the slow-roll approximation. This has been achieved by calculating the power spectra up to the second-order in the slow-roll approximation by the Green's function method \cite{Stewart2001PLB, Gong2004CQG}. Then, the spectra at the second-order in the slow-roll approximation have been re-derived by using the WKB approximation \cite{Martin-WKB2003PRD,CasadioWKB2005PLB}, improved-WKB approximation \cite{ImprovedWKB2005PRD}, and also the first-order uniform 
asymptotic approximation \cite{Martin2013JCAP,Temme}. Except the above mentioned approximations, other methods are also available for obtaining the power spectra and spectral indices, such as the Bessel function approximation \cite{Schwarz2001PLB}, and the phase integral method \cite{Rojas2007PRD}.

The uniform asymptotic approximation method mentioned above was first applied to inflationary cosmology in the framework of general relativity in Refs. \cite{uniformPRL, uniformPRD}, and then was extended by us \cite{ZhuUniform2014PRD1} to the more general case where the dispersion relation is not necessarily linear, and in general has multi- and/or high-order turning points. Furthermore, by considering the more accurate high-order uniform asymptotic  approximations, we have obtained the general expressions of the power spectra and spectral indices up to the third-order in the uniform asymptotic approximation, at which the error bounds are $\lesssim 0.15\%$ \cite{ZhuUniform2014-2}. With these results we have also calculated explicitly the power spectra and spectral indices of scalar and tensor perturbations with a modified dispersion relation up to the second-order in the slow-roll approximation. These results have been checked by comparing them with those obtained by the Green's function method, and are shown to be accurate enough to match the requirements of current and forthcoming observations.

The success of the high-order uniform asymptotic approximations motivates us to apply this powerful technique to other inflationary scenarios. In particular, in this paper we shall apply it to the most general single field inflation, the $k$-inflation. The most distinguishable feature of the $k$-inflation is that the scalar perturbations obey an equation of motion with a time-dependent sound speed. This makes it very difficult to calculate the corresponding power spectra and spectral indices, although with some additional assumptions, the power spectra of the $k$-inflation can be obtained by using the Green's function method \cite{Wei2004PLB}. For a general sound speed, after introducing a simple hierarchy of parameters, related to the sound speed of the  scalar perturbations and its successive derivatives, the authors in Refs. \cite{Martin2013JCAP, Martin2008PRD} have worked out explicitly the power spectra and spectral indices by using the first-order uniform asymptotic approximation, at which the error bounds in general are $\lesssim 15\%$, although further improvement can be achieved, similar to that done in the relativistic case \cite{uniformPRD}.

However, to match with the accuracy of the current and forthcoming observations, as pointed out in \cite{ZhuUniform2014-2}, consideration of the high-order corrections in the uniform asymptotic approximation is highly demanded. In this paper, with the general expressions of power spectra and spectral indices we obtained in \cite{ZhuUniform2014-2}, we calculate explicitly the power spectra and spectral indices of scalar and tensor perturbations of the $k$-inflation up to the third-order in the uniform asymptotic approximation. These expressions represent a significant improvement of the previous results obtained by other methods. Furthermore, by comparing our expressions with the ones obtained by other methods, such as the first-order uniform asymptotic approximation, the Green's function method, WKB approximation, and improved WKB approximation, we show explicitly the relative errors among the results obtained by these different methods.

The paper is organized as follows. In Sec. II, we present a brief review of the $k$-inflation, and in Sec. III,  we give the most general formulas of the high-order uniform asymptotic approximations. Then in Sec. IV, with these general expressions we calculate explicitly the power spectra, spectral indices, and running of the spectral indices of both scalar and tensor perturbations in the slow-roll $k$-inflation. In Sec. V, we present a detailed comparison of the results with the ones obtained by other methods. Our main conclusions are summarized in Sec. VI.

Before proceeding further, we note that it must not be confused with the order of {\em the uniform asymptotic approximations} and the order of {\em the slow-roll parameters}.  The
 former is defined  by the parameter $\lambda$, appearing in Eq.(\ref{eom}), while the latter   is characterized by the parameters $\epsilon_n$ and $q_n$, defined, respectively, in
 Eqs.(\ref{Hubbleflow}) and (\ref{sound flow}). These represent two independent sets of parameters. As a result,  their expansions are also independent one from the other. In addition, 
  in this paper we do not consider the  pivot expansion.

\section{Scalar and tensor perturbations of the $k$-inflation}
\renewcommand{\theequation}{2.\arabic{equation}} \setcounter{equation}{0}

In this section, we present a brief introduction of the scalar and tensor perturbations of the $k$-inflation. In general, the action of the $k$-inflation can be written in the form,
\bqn
S=\frac{1}{2}\int d^4 x \sqrt{-g} \left[R+2 P(X,\phi)\right],
\eqn
where $g$ is the determinant of the metric, $R$ is the $4$D Ricci scalar, $\phi$ denotes a scalar field and
\bqn
X = \frac{1}{2} g^{\mu\nu} \partial_\mu \phi \partial_\nu \phi,
\eqn
is the kinetic term. To be stable, the $k$-inflation must satisfy the following two conditions \cite{Martin2008PRD},
\bqn
\frac{\partial P}{\partial X}>0,\;\;\;\;2 X \frac{\partial^2 P}{\partial X^2}+\frac{\partial P}{\partial X}>0.
\eqn

Let us consider a flat universe for simplicity, for which the background metric is
\bqn
ds^2&=&-dt^2+a^2(t) (dx^2+dy^2+dz^2)\nb\\
&=& a^2(\eta) (-d\eta^2+dx^2+dy^2+dz^2),
\eqn
where $a$ is the scale factor of the universe,  and $\eta$ is the conformal time defined as $d\eta=dt/a$. For the background evolution during the inflation, it is convenient to define a hierarchy of Hubble flow parameters,
\bqn\lb{Hubbleflow}
\epsilon_{n+1}\equiv \frac{d\ln\epsilon_n}{d\ln a}, \;\;\;\;\;\;\epsilon_0\equiv \frac{H_{\text{ini}}}{H},
\eqn
where $H=\dot a/a$ is the Hubble parameter, and a dot denotes derivative with respect to the cosmic time $t$.

In general, the perturbations produced during the inflationary epoch are governed by the master equation
\bqn
\mu''_k(\eta)+\left(c_s^2(\eta)k^2-\frac{z''}{z}\right)\mu_k(\eta)=0,
\eqn
where $\mu_k(\eta)$ denotes the inflationary mode function,
a prime denotes differentiation with respect to the conformal time $\eta$, $k$ is the co-moving wavenumber, $c_s(\eta)$ and $z(\eta)$ depend on the background and the types of the perturbations (scalar and tensor).

For scalar perturbations, we have
\bqn
c_s^2(\eta) \equiv \frac{P_{,X}}{P_{,X}+2X P_{,XX}},
\eqn
where a subscript ``, X" represents differentiation with respect to $X$. Similar to the definitions of the hierarchy of Hubble parameters, the authors in Refs.\cite{Martin2013JCAP, Martin2008PRD} introduced a hierarchy of the sound speed, which are given by
\bqn\lb{sound flow}
q_{n+1} \equiv \frac{d\ln q_n}{d\ln a},\;\;\;\;\;q_0 \equiv \frac{c_{\text{ini}}}{c_s}.
\eqn
In general the parameters are assumed to be small, i.e., $q_n \ll 1$. With the above definitions we have
\bqn
\frac{z''(\eta)}{z(\eta)} &=& a^2 H^2
\Big(2-\epsilon_1+\frac{3}{2}\epsilon_2+3 q_1+\frac{1}{4} \epsilon_2^2-\frac{1}{2} \epsilon_1 \epsilon_2\nb\\
&&~~~~~~~~~+\frac{1}{2} \epsilon_2\epsilon_3-\epsilon_1 q_1+\epsilon_2 q_1+q_1^2+q_1 q_2\Big),\nb\\
\eqn
in which $z(\eta)=\sqrt{2 \epsilon_1} a(\eta)/c_s(\eta)$. For tensor perturbations, the corresponding expressions become simpler. In particular, we have $c_s(\eta)=1$ and
\bqn
\frac{z''(\eta)}{z(\eta)} = a^2 H^2 (2-\epsilon_1),
\eqn
where $z(\eta)=a(\eta)$.

\section{Power spectra and spectral indices in the uniform asymptotic approximation}
\renewcommand{\theequation}{3.\arabic{equation}} \setcounter{equation}{0}

\subsection{Brief introduction of the uniform asymptotic approximation}

In this section, we first present a brief introduction to the {\em uniform asymptotic approximation method} with high-order corrections. Most of the expressions and results presented here can be found in Refs. \cite{uniformPRD, ZhuUniform2014PRD1,ZhuUniform2014-2}.

Following Refs.~\cite{Olver1974,ZhuUniform2014-2}, by introducing a dimensionless variable $y=-k\eta$, let us first write the equation of the mode function in the form,
\bqn\lb{eom}
\frac{d^2\mu_k(y)}{dy^2}=\left[\lambda^2\hat g(y)+q(y)\right]\mu_k(y).
\eqn
In the above the parameter $\lambda$ is used to trace the order of the uniform approximations, and $\lambda^2\hat g(y)=g(y)$. Usually $\lambda$ is supposed to be large, and can be absorbed into $g(y)$. Thus, when we turn to the final results, we can set $\lambda=1$ for the sake of simplification. Then, it is easy to show that
\bqn
\lambda^2\hat g(y)+q(y) \equiv - \frac{1}{k^2} \left(c^2_s(\eta)k^2-\frac{z''(\eta)}{z(\eta)}\right).
\eqn
In general, $\hat g(y)$ and $q(y)$ have two poles (singularities): one is at $y=0^+$ and the other is at $y=+\infty$. As we discussed in \cite{ZhuUniform2014PRD1} (see also \cite{uniformPRL,Olver1974}), if these two poles are both second-order or higher, one must choose
\bq
\lb{qF}
q(y)=- \frac{1}{4y^2},
\eq
for the convergence of the error control functions. In addition, the function $\hat g(y)$ can vanish at various points, which are called turning points or zeros, and the approximate solution of the mode function $\mu_k(y)$ depends on the behavior of $\hat g(y)$ and $q(y)$ near these turning points.

To proceed further, let us first introduce the Liouville transformations with two new variables $U(\xi)$ and $\xi$ via the relations
\bqn
\lb{Olver trans}
U(\xi)&=& \chi^{1/4} \mu_k(y),\;\;\; \xi'^2 =  \frac{|\hat g(y)|}{f^{(1)}(\xi)^2},
\eqn
where $ \chi \equiv \xi'^2,\; \xi'=d\xi/dy$, and
\bqn
\lb{OlverTransB}
f(\xi)&=& \int^y \sqrt{|\hat g(y)|} dy,\;\;\;  f^{(1)}(\xi)=\frac{df(\xi)}{d\xi}.
\eqn
Note that $\chi$ must be regular and not vanish in the intervals of interest. Consequently, $f(\xi)$ must be chosen so that $f^{(1)}(\xi)$ has zeros and singularities of the same type as that of $\hat g(y)$. As shown in \cite{ZhuUniform2014PRD1, ZhuUniform2014-2}, such a requirement plays an
essential role in determining the approximate solutions. In terms of $U$ and $\xi$, Eq. (\ref{eom}) takes the form
\bqn\lb{eomU}
\frac{d^2 U}{d\xi^2}&=&\left[\pm \lambda^2 f^{(1)}(\xi)^2+\psi(\xi)\right]U,
\eqn
where
\bqn\lb{psi}
\psi(\xi)=\frac{q(y)}{\chi}-\chi^{-3/4} \frac{d^2(\chi^{-1/4})}{dy^2},
\eqn
and the signs ``$\pm$" correspond to $\hat g(y)>0$ and $\hat g(y)<0$, respectively. Considering $\psi(\xi) =0$ as the first-order approximation, one can choose $f^{(1)}(\xi)$ so that the first-order approximation can be as close to the exact solution as possible with the guidelines of minimizing  the error functions constructed below, and   solving it in terms of known functions. Clearly, such a choice sensitively depends on the behavior of the functions $\hat g(y)$ and $q(y)$ near the poles and turning points.

For the case in which $\hat g(y)$ has only one single turning point, we can choose
\bqn
f^{(1)}(\xi)=\pm \xi,
\eqn
here $\xi=\xi(y)$ is a monotone decreasing function, and $\pm$ correspond to $\hat g(y)\geq 0$ and $\hat g(y) \leq 0$, respectively. Following Olver \cite{Olver1974}, the general solution of Eq.~(\ref{eomU}) can be written as
\bqn\lb{appro}
U(\xi)&=&\alpha_0 \Bigg[\text{Ai}(\lambda^{2/3} \xi) \sum_{s=0}^{n} \frac{A_s(\xi)}{\lambda^{2s}}\nb\\
&&~~~~~~~+\frac{\text{Ai}'(\lambda^{2/3}\xi)}{\lambda^{4/3}} \sum_{s=0}^{n-1} \frac{B_s(\xi)}{\lambda^{2s}}+\epsilon_3^{(2n+1)}\Bigg]\nb\\
&&+\beta_0 \Bigg[\text{Bi}(\lambda^{2/3} \xi) \sum_{s=0}^{n} \frac{A_s(\xi)}{\lambda^{2s}}\nb\\
&&~~~~~~~~+\frac{\text{Bi}'(\lambda^{2/3}\xi)}{\lambda^{4/3}} \sum_{s=0}^{n-1} \frac{B_s(\xi)}{\lambda^{2s}}+\epsilon_4^{(2n+1)}\Bigg],\nb\\
\eqn
where $\text{Ai}(x)$ and $\text{Bi}(x)$ represent the Airy  functions, $\epsilon_3^{(2n+1)}$ and $\epsilon_{4}^{(2n+1)}$ are errors of the approximate solution, and
\bqn\lb{AB}
&& A_0(\xi)=1,\;\;\nb\\
&&B_s=\frac{\pm 1}{2 (\pm \xi)^{1/2}}\int_0^\xi \{\psi(v) A_s(v)-A''_s(v)\}\frac{dv}{(\pm v)^{1/2}},\nb\\
&&A_{s+1}(\xi)=-\frac{1}{2} B'_s(\xi)+\frac{1}{2} \int \psi(v) B_s(v) dv,\nb\\
\eqn
where $\pm$ correspond to $\xi \geq 0$ and $\xi\leq 0$, respectively. The up bounds of $\epsilon_3^{(2n+1)}$ and $\epsilon_4^{(2n+1)}$ can be expressed as
\bqn\lb{error}
&&\frac{\epsilon_{3}^{(2n+1)}}{M(\lambda^{2/3} \xi)}, \;\;\frac{\partial \epsilon_{3}^{(2n+1)}/\partial\xi}{\lambda^{2/3} N(\lambda^{2/3}\xi)}\nb\\
&& ~~~~~\leq 2 E^{-1}(\lambda^{2/3}\xi)  \exp{\left[\frac{2\kappa_0 \mathscr{V}_{\alpha,\xi}(|\xi^{1/2}|B_0)}{\lambda}\right]} \nb\\
&&~~~~~~~~~~~\times \frac{\mathscr{V}_{\alpha,\xi}(|\xi^{1/2}|B_n)}{\lambda^{2n+1}},\nb\\
&&\frac{\epsilon_{4}^{(2n+1)}}{M(\lambda^{2/3} \xi)}, \;\;\frac{\partial \epsilon_{4}^{(2n+1)}/\partial\xi}{\lambda^{2/3} N(\lambda^{2/3}\xi)} \nb\\
&&~~~~~\leq 2 E(\lambda^{2/3}\xi) \exp{\left[\frac{2\kappa_0 \mathscr{V}_{\xi,\beta}(|\xi^{1/2}|B_0)}{\lambda}\right]} \nb\\
&&~~~~~~~~~~~\times\frac{\mathscr{V}_{\xi,\beta}(|\xi^{1/2}|B_n)}{\lambda^{2n+1}},
\eqn
where the definitions of $M(x)$, $N(x)$, $\kappa_0$, and $\mathscr{V}_{a,b}(x)$ can be found in \cite{ZhuUniform2014PRD1}.

\subsection{Power spectra and spectral indices up to the third-order in the uniform asymptotic approximation}

With the approximate solution given above, now let us calculate the power spectra and spectral indices from the approximate solution. We assume that the universe was initially at the adiabatic vacuum,
\bqn
\lim_{y\to +\infty} \frac{1}{\sqrt{2 \omega_k(\eta)}} e^{-i \int \omega_k(\eta) d\eta}.
\eqn
Then, we need to match this initial state with the approximate solution (\ref{appro}).  However, the approximate solution involves many high-order terms, which are  complicated and not easy to handle. In order to simplify the calculations, we first study their behavior in the limit $y\rightarrow +\infty$. Let us start with the $B_0(\xi)$ term in Eq.(\ref{AB}), which satisfies
\bqn
B_0(\xi)=-\frac{1}{2\sqrt{-\xi}} \int_{0}^{\xi} \frac{\psi(v)}{\sqrt{-v}}dv=-\frac{\mathscr{H}(\xi)}{2\sqrt{-\xi}},
\eqn
where $\mathscr{H}(\xi)\equiv \int_{0}^{\xi} dv \psi(v)/|v|^{1/2}$ is the associated error control function of the approximate solution (\ref{appro}), and in the above we had used $A_0(\xi)=1$. The error control function $\mathscr{H}(\xi)$ is well behaved around the turning point $\bar y_0$ and converges  when $y\to +\infty$. As a result, we have
\bqn
\lim_{y\rightarrow +\infty} B_0(\xi) =-\frac{\mathscr{H}(-\infty)}{2\sqrt{-\xi}}.
\eqn
Then, let us turn to $A_1$, which is
\bqn
A_1(\xi)=-\frac{1}{2} B_0'(\xi)+\frac{1}{2}\int_{0}^{\xi} \psi(v)B_0(v)dv.
\eqn
In the limit $y\rightarrow +\infty$, $B_0'(\xi)$ vanishes, and we find
\bqn
\lim_{y\to+\infty} A_1(\xi) &=&-\frac{1}{2} \int_{0}^{\xi} \frac{\psi(v)}{\sqrt{-v}} \left[\frac{1}{2}\int_{0}^{v} \frac{\psi(u)}{\sqrt{-u}} du\right] dv\nb\\
&=&-\frac{1}{2} \left[\frac{\mathscr{H}(-\infty)}{2}\right]^2.
\eqn
Note that in the above we had used the formula
\bqn
&&n! \int_{\xi_0}^{\xi} f(\xi_n) \int_{\xi_0}^{\xi_n} f(\xi_{n-1})\cdots \int_{\xi_0}^{\xi_2}f(\xi_1)d\xi_1d\xi_2\cdots d\xi_n\nb\\
&&~~~~~~~~~~~~~~~~~~~~~~~~~~~~~~=\left[\int_{\xi_0}^{\xi} f(v)dv\right]^n.
\eqn
Thus, up to the third-order in the uniform asymptotic approximation, we obtain
\bqn
A_0(\xi)+\frac{A_1(\xi)}{\lambda^2} &=&1-\frac{1}{2\lambda^2} \left[\frac{\mathscr{H}(-\infty)}{2}\right]^2+\mathcal{O}\left(\frac{1}{\lambda^3}\right),\;\;\;\;\;\nb\\
\frac{B_0(\xi)}{\lambda}&=& -\frac{1}{\sqrt{-\xi}} \frac{\mathscr{H}(-\infty)}{2\lambda}+\mathcal{O}\left(\frac{1}{\lambda^3}\right). ~~~
\eqn
Then, using the asymptotic form of Airy  functions in the limit $\xi\to -\infty$, and comparing the solution $\mu_k(y)$ with the initial state, we obtain
\bqn
\alpha_0=\sqrt{\frac{\pi}{2k}} \frac{1}{(A_0+A_1/\lambda^2)-i \sqrt{-\xi} B_0/\lambda},\nb\\
\beta_0=i \sqrt{\frac{\pi}{2k}} \frac{1}{(A_0+A_1/\lambda^2)-i \sqrt{-\xi} B_0/\lambda},
\eqn
where we have
\bqn
(A_0+A_1/\lambda^2)-i \sqrt{-\xi} B_0/\lambda =(1+\mathcal{O}(1/\lambda^3)) e^{i\theta}. ~~~~
\eqn
Here $\theta$ is an irrelevant phase factor, and without loss of the generality, we can set $\theta=0$. Thus, we  finally get
\bqn
\alpha_0&\simeq& \sqrt{\frac{\pi}{2k}} \left(1+\mathcal{O}(1/\lambda^3)\right),\nb\\
\beta_0&\simeq & i\sqrt{\frac{\pi}{2k}} \left(1+\mathcal{O}(1/\lambda^3)\right).
\eqn

After determining the coefficients $\alpha_0$ and $\beta_0$, we can calculate the power spectra of the perturbations. As $y \rightarrow 0$, only  the growing mode is relevant, thus we have
\bqn
\mu_k(y)&\simeq& \beta_0 \left(\frac{\xi}{\hat{g}(y)}\right)^{1/4}\Bigg[\text{Bi}(\lambda^{2/3}\xi)\sum_{s=0}^{+\infty} \frac{B_s(\xi)}{\lambda^{2s}}\nb\\
&&\;\;\;\;\;\;\;\;\;\;\;\;+\frac{\lambda^{2/3}\text{Bi}'(\lambda^{2/3}\xi)}{\lambda^2}\sum_{s=0}^{+\infty}\frac{B_s(\xi)}{\lambda^{2s}}\Bigg].\;\;\;\;\;\;\;\;
\eqn
In order to calculate  the power spectra to high-order in terms of the uniform asymptotic approximation parameter $\lambda$, let us first consider the $B_0(\xi)$ term, which satisfies
\bqn
\lim_{y\to 0 }B_0(\xi)=\frac{1}{2\xi^{1/2}} \int_{0}^{\xi} \frac{\psi(v)}{v^{1/2}}dv=\frac{\mathscr{H}(+\infty)}{2\xi^{1/2}}.
\eqn
In the above we had used the relation $\xi^{1/2}d\xi=-\sqrt{\hat{g}}dy$.
Based upon the $B_0$ term, we can get the $A_1$ term, which is
\bqn
\lim_{y\to 0} A_1(\xi)&=&\frac{1}{4} \int_{0}^{\xi} \frac{\psi(v)}{v^{1/2}}  \int_{0}^{v} \frac{\psi(u)}{u^{1/2}} du dv\nb\\
&=&\frac{1}{2} \left[\frac{\mathscr{H}(+\infty)}{2}\right]^2.
\eqn
Then, up to the third order in the uniform asymptotic approximation, and considering the asymptotic forms of the Airy functions in the limit $\xi\rightarrow +\infty$, we find
\bqn
\lim_{y\rightarrow 0} \mu_k(y)&=& \frac{\beta_0 e^{\frac{2}{3}\lambda \xi^{2/3} } }{ \lambda^{1/6} \hat{g}^{1/4} \pi^{1/2} }
\Bigg[1+\frac{\mathscr{H}(+\infty)}{2\lambda}+\frac{\mathscr{H}(+\infty)^2}{8\lambda^2}\nb\\
&&~~~~~~~~~~~~~~~~~~~~~~~~+\mathcal{O}(1/\lambda^3)\Bigg].
\eqn
Hence, the power spectra are given by
\bqn\lb{pw}
\Delta^2(k)&\equiv& \frac{k^3}{2\pi^2} \left|\frac{\mu_k(y)}{z}\right|^2_{y\to 0^{+}}\nb\\
&\simeq&\frac{k^2}{4\pi^2}\frac{-k\eta}{z^2(\eta) \nu(\eta)}\exp\left(2\int_y^{\bar y_0}\sqrt{\hat{g}(\hat{y})}d\hat{y}\right)\nb\\
&&\;\times \left[1+\frac{\mathscr{H}(+\infty)}{\lambda}+\frac{\mathscr{H}^2(+\infty)}{2\lambda^2}+\mathcal{O}(1/\lambda^3)\right].\nb\\
\eqn

From the power spectra presented above, one can obtain the general expression of the spectral indices, which now is given by
\bqn\lb{spectral indices}
n-1&\equiv&\frac{d\ln \Delta^2(k)}{d\ln k}\nb\\
&\simeq&3+2\frac{d}{d\ln k} \int_y^{\bar y_0} \sqrt{\hat g(\hat y)} d\hat y+\frac{1}{\lambda}\frac{d\mathscr{H}(+\infty)}{d\ln k}\nb\\
&&+\mathcal{O}\left(\frac{1}{\lambda^3}\right).
\eqn
Note that the third term in the above expression contains contributions of both the second- and third-order corrections in the uniform asymptotic approximations.

\section{Applications to the  $k$-inflation}
\renewcommand{\theequation}{4.\arabic{equation}} \setcounter{equation}{0}

Now we arrive at the position to calculate the power spectra and spectral indices of the k-inflation from the general formulas (\ref{pw}) and (\ref{spectral indices}). In order to do so, we need to perform the integral of $\sqrt{g(y)}$ in Eq.(\ref{pw}) and calculate the error control function $\mathscr{H}(+\infty)$.  However, this becomes very complicated,  if the explicit form of $\nu(\eta)$ and $c_s(\eta)$ is not specified. In this paper, we shall focus on the slow-roll $k$-inflation and consider the slow-roll expansions of the power spectra at second-order and corresponding spectral indices at the third-order in terms of the slow-roll parameters, for both scalar and tensor perturbations. As we discussed in \cite{ZhuUniform2014-2}, in the slow-roll inflation, it is convenient to consider the following expansions,
\bqn
\nu(\eta)\simeq \bar \nu_0+\bar \nu_1\ln\frac{y}{\bar y_0}+\frac{1}{2}\bar \nu_2\ln^2\frac{y}{\bar y_0},
\eqn
with
\bqn
\bar \nu_1 \equiv \frac{d\nu(\eta)}{d\ln(-\eta)}\Bigg|_{\eta_0},\;\;\;\bar \nu_2\equiv \frac{d^2\nu(\eta)}{d\ln^2(-\eta)}\Bigg|_{\eta_0},
\eqn
and
\bqn
c_s(\eta)\simeq \bar c_0+\bar c_1\ln\frac{y}{\bar y_0}+\frac{1}{2}\bar c_2\ln^2\frac{y}{\bar y_0},
\eqn
with
\bqn
\bar c_1 \equiv \frac{dc_s(\eta)}{d\ln(-\eta)}\Bigg|_{\eta_0},\;\;\;\bar c_2\equiv \frac{d^2c_s(\eta)}{d\ln^2(-\eta)}\Bigg|_{\eta_0}.
\eqn
In the slow-roll $k$-inflation, we have $\nu(\eta)\simeq \frac{3}{2}+\mathcal{O}(\epsilon)$, $\bar \nu_1\simeq \mathcal{O}(\epsilon^2)$, $\bar \nu_2 \simeq \mathcal{O}(\epsilon^3)$, and $\bar c_0 \simeq \mathcal{O}(1), \bar c_2 \simeq\mathcal{O}(\epsilon), \;\;\bar c_2 \simeq \mathcal{O}(\epsilon^2)$, where $\epsilon$ represents the slow-roll parameters and $\mathcal{O}(\epsilon^n)$ denotes the $n$-th order in the slow-roll approximations. The slow-roll expansions of all the above quantities can be found in Appendix A.

With the above expansions, we notice that
\bqn
\sqrt{g(y)}
&\simeq& \frac{\sqrt{\bar \nu_0^2-\bar c_0^2 y^2}}{y}+\frac{\bar \nu_0 \bar \nu_1-\bar c_0 \bar c_1 y^2}{y \sqrt{\bar \nu_0^2-\bar c_0^2 y^2}}\ln\frac{y}{\bar y_0}\nb\\
&&+\Bigg(\frac{\bar \nu_0 \bar \nu_2 }{2y \sqrt{\bar \nu_0^2-\bar c_0^2 y^2}}-\frac{\bar c_0 \bar c_2 y}{2 \sqrt{\bar \nu_0^2-\bar c_0^2 y^2}}\nb\\
&&\;\;\;\;\;\;\;\;\;-\frac{(\bar c_0 \bar \nu_1+\bar \nu_0 \bar c_1)^2y}{2 (\bar \nu_0^2-\bar c_0^2 y^2)^{3/2}}\Bigg)\ln^2\frac{y}{\bar y_0}.
\eqn
Therefore, the integral $\int \sqrt{\hat g}dy$ can be divided into three parts,
\bqn\lb{g3parts}
\int_{y}^{\bar y_0} \sqrt{\hat g(\hat y)}dy=I_1+I_2+I_3,
\eqn
where
\bqn
\lim_{y\to 0}I_1&=& -\bar \nu_0 \left(1+\ln \frac{y}{2\bar y_0}\right),\nb\\
\lim_{y\to 0}I_2&=&\frac{(1-\ln2) \bar c_1 \bar \nu_0}{\bar c_0}-\left(\frac{ \pi^2}{24}-\frac{\ln^22}{2}+\frac{1}{2}\ln^2\frac{y}{\bar y_0}\right)\bar \nu_1,\nb\\
\lim_{y\to 0}I_3&=&- \bar \nu_0 \left(\frac{\pi^2-12\ln^22}{24}\right)\frac{\bar c_1^2}{\bar c_0^2}\nb\\
&&-\bar \nu_0 \left(1-\frac{\pi^2}{24}-\ln2+\frac{\ln^22}{2}\right)\frac{\bar c_2}{\bar c_0}\nb\\
&&+\left(\frac{\zeta(3)}{4}-\frac{\pi^2\ln2}{24}+\frac{\ln^32}{6}-\frac{1}{6}\ln^3\frac{y}{\bar y_0}\right)\bar \nu_2.\nb\\
\eqn
Note that in the above we keep the $\bar \nu_2 \sim \mathcal{O}(\epsilon^3)$ term in the expression of $I_3$, this is because when we calculate the spectral indices from Eq. (\ref{spectral indices}), the $k$-derivative of this term provides cancellation of the $\ln^2(y/\bar y_0)$ divergence of the $k$-derivative of $\bar \nu_1$ term in $I_2$.

Now, we turn to consider the error control function $\mathscr{H}$. After some lengthy calculations we arrive at
\bqn
\mathscr{H}(\xi)&=&\frac{5}{36} \left\{\int_{\bar y_0}^{\tilde{y}}\sqrt{\hat{g}(\tilde{y})}d\tilde{y}\right\}^{-1}\Bigg|^{y}_{\bar y_0}\nb\\
&&-\int_{\bar y_0}^{y} \left\{\frac{q}{\hat{g}}-\frac{5\hat{g}'^2}{16\hat{g}^3}+\frac{\hat{g}''}{4\hat{g}^2}\right\}\sqrt{\hat{g}}dy.
\eqn
In the limit $y\to 0$, the above expression can be casted in the form
\bqn\lb{Hinfty}
\mathscr{H}(+\infty)&\simeq& \frac{1}{6\bar \nu_0} \left(1+\frac{\bar c_1}{\bar c_0}\right)-\frac{\bar\nu_1(23+12\ln 2)}{72\bar \nu_0^2}\nb\\
&&+\frac{37 \bar c_1^2}{36 \bar c_0^2 \bar \nu _0}-\frac{5 \bar c_2}{36 \bar c_0 \bar \nu _0}-\frac{17 \bar c_1^2 \ln2}{70 \bar c_0^2 \bar \nu _0}+\frac{\bar c_2 \ln2}{6 \bar c_0 \bar \nu _0}.\nb\\
\eqn
Once we get the integral of $\sqrt{g(y)}$ in Eq.~(\ref{g3parts}) and error control function in Eq.~(\ref{Hinfty}), from Eq.~(\ref{pw}) one can easily calculate the power spectra.

Now we turn to consider the corresponding spectral indices. In order to do this, we  first specify the $k$-dependence of $\bar\nu_0(\eta_0)$, $\bar \nu_1(\eta_0)$ through  $\eta_0 = \eta_0(k)$. From the relation $-k\eta_0=\bar \nu_0(\eta_0)/\bar c_0(\eta_0)$, after lengthy technical calculations, we find
\bqn
\frac{d\ln(-\eta_0)}{d\ln k}& \simeq& -1+\frac{\bar c_1}{\bar c_0}-\frac{\bar\nu_1}{\bar\nu_0}-\left(\frac{\bar c_1}{\bar c_0}-\frac{\bar\nu_1}{\bar\nu_0}\right)^2\nb\\
&&+\left(\frac{\bar c_1}{\bar c_0}-\frac{\bar\nu_1}{\bar\nu_0}\right)^3.
\eqn
\begin{widetext}
Then, the spectral indices are given by
\bqn
n-1&\simeq&
\left(3-2 \bar \nu _0\right)
+\frac{2 \bar c_1 \bar \nu_0}{\bar c_0}
+\left(\frac{1}{6\bar \nu_0^2}-2\ln2\right)\bar \nu_1
+\left(\frac{2\bar  \nu _0 \ln2}{\bar c_0}-\frac{2 \bar \nu _0}{\bar c_0}-\frac{1}{6 \bar c_0 \bar \nu _0}\right)\bar c_2
+\left(\frac{23+12\ln2}{72 \bar \nu _0^2}+\frac{\pi ^2}{12}-\ln^22\right)\bar \nu_2
\nb\\&&
+\left(\frac{1-12 \bar \nu _0^2 \ln2}{6 \bar c_0^2 \bar \nu _0}\right)\bar c_1^2
+\frac{4\bar \nu_1\bar c_1\ln2}{\bar c_0}
+\frac{5-6\ln2}{36\bar c_0\bar \nu_0}\bar c_3
+\left(2-\frac{\pi ^2}{12}+\ln^22-\ln4\right)\frac{\bar \nu_0\bar c_3}{\bar c_0}
+\left(\frac{1}{9 }-\frac{\ln2}{35}\right)\frac{17\bar c_1^3}{ \bar c_0^3 \bar \nu _0}
\nb\\&&
+\left(2\ln2+2\ln^22-\frac{\pi^2}{6}\right)\frac{\bar \nu_0 \bar c_1^3}{\bar c_0^3 }
+\left(\frac{\pi ^2 \bar\nu _0}{4 \bar c_0^2}
-\frac{73}{36 \bar c_0^2 \bar \nu _0}
-\frac{3 \bar \nu _0 \ln^22}{\bar c_0^2}
+\frac{137 \ln2}{210 \bar c_0^2 \bar \nu _0}\right)\bar c_1\bar c_2.
\eqn

Similarly, after some tedious calculations, we find that the running of the spectral index $\alpha_s \equiv dn_s/d\ln k$ is given by
\bqn
\alpha(k)
&\simeq&
\left(\frac{6\bar \nu _0 \ln^22}{\bar c_0^4}-\frac{51 \ln2}{35 \bar c_0^4 \bar \nu _0}+\frac{5\bar \nu _0\ln4 }{\bar c_0^4}+\frac{2 \bar\nu _0}{\bar c_0^4}-\frac{\pi ^2 \bar \nu _0}{2 \bar c_0^4}+\frac{16}{3 \bar c_0^4 \bar \nu _0}\right) \bar c_1^4
+\left(\frac{1}{3 \bar c_0^3 \bar \nu _0}-\frac{2 \bar \nu _0\ln4 }{\bar c_0^3}-\frac{2 \bar \nu _0}{\bar c_0^3}\right) \bar c_1^3
\nb\\&&
+\left(\frac{58 \ln 2}{21 \bar c_0^3 \bar \nu _0}+\frac{\pi ^2 \bar \nu _0}{\bar c_0^3}-\frac{83}{9 \bar c_0^3 \bar \nu _0}-\frac{12 \bar \nu _0 \ln^22}{\bar c_0^3}-\frac{12 \bar \nu _0 \ln2}{\bar c_0^3}\right) \bar c_2 \bar c_1^2
+\frac{2 \bar \nu _0 \bar c_1^2}{\bar c_0^2}
+\left(\frac{6 \ln2}{\bar c_0^2}+\frac{6}{\bar c_0^2}+\frac{1}{6 \bar c_0^2 \bar \nu _0^2}\right) \bar \nu _1 \bar c_1^2
\nb\\&&
+\left(\frac{6 \bar \nu _0\ln2 }{\bar c_0^2}-\frac{1}{2 \bar c_0^2 \bar \nu _0}\right) \bar c_2 \bar c_1
+\left(\frac{4 \bar \nu _0 \ln^22}{\bar c_0^2}-\frac{86 \ln2}{105 \bar c_0^2 \bar \nu _0}-\frac{\pi ^2 \bar \nu _0}{3 \bar c_0^2}+\frac{2}{\bar c_0^2 \bar \nu _0}\right) \bar c_3 \bar c_1
+\left(\frac{1}{6 \bar c_0 \bar \nu _0^2}-\frac{6 \ln2}{\bar c_0}\right) \bar \nu _2 \bar c_1
\nb\\&&
-\frac{4 \bar \nu _1 \bar c_1}{\bar c_0}
+\left(\frac{3 \bar \nu _0 \ln^22}{\bar c_0^2}-\frac{137 \ln2}{210 \bar c_0^2 \bar \nu _0}-\frac{\pi ^2 \bar \nu _0}{4 \bar c_0^2}+\frac{73}{36 \bar c_0^2 \bar \nu _0}\right) \bar c_2^2
+\left(\frac{1}{3 \bar \nu _0^3}+\frac{2}{\bar \nu _0}\right) \bar \nu _1^2
+\left(\frac{2 \bar \nu _0}{\bar c_0}+\frac{1}{6 \bar c_0 \bar \nu _0}-\frac{2\bar \nu _0 \ln2 }{\bar c_0}\right) \bar c_3
\nb\\&&
+\frac{\pi ^2 \bar c_4 \bar \nu _0}{12 \bar c_0}
+\left(-\frac{6 \ln2}{\bar c_0}-\frac{1}{6 \bar c_0 \bar \nu _0^2}\right)\bar c_2 \bar \nu_1
+2 \bar \nu _1
+\left(\ln4-\frac{1}{6 \bar \nu _0^2}\right) \bar \nu _2
+\left(\ln^22-\frac{\pi ^2}{12}-\frac{\ln2}{6 \bar \nu _0^2}-\frac{23}{72 \bar \nu _0^2}\right) \bar \nu _3
\nb\\&&
+\frac{\bar c_4 \bar \nu _0\ln4 }{\bar c_0}
-\frac{2 \bar c_2 \bar \nu _0}{\bar c_0}-\frac{\bar \nu _0\bar c_4 \ln^22 }{\bar c_0}-\frac{2 \bar c_4 \bar \nu _0}{\bar c_0}-\frac{5 \bar c_4}{36 \bar c_0 \bar \nu _0}+\frac{\ln2 \bar c_4}{6 \bar c_0 \bar \nu _0}.
\eqn

\subsection{Scalar Perturbations}

With the above expressions and the slow-roll expansions of $c_1,\;c_2,\;c_3,\;\nu_0,\;\nu_1,\;\nu_2,\;\text{and}\;\nu_3$, which are presented in Appendix B, we find that, up to second-order in the slow-roll approximation, the scalar spectrum can be cast in the form,
\bqn
\Delta^2_s(k) &\simeq&
\frac{181\bar H ^2}{72 e^3\pi^2 \bar \epsilon_1 \bar c_0}
\Bigg\{
1
+ \left(\frac{429}{181}-\ln2\right)\bar q_1
+ \left(\ln4-\frac{496}{181}\right)\bar \epsilon _1
+ \left(\ln2-\frac{67}{181}\right)\bar \epsilon _2
+\left(\frac{2095}{1086}+\frac{\ln^22}{2} -\frac{600 \ln8}{1267}\right)\bar q_1^2
\nb\\&&
~~
+\left(\frac{4865}{1629}-\frac{\pi ^2}{24}+\frac{\ln^22}{2}-\frac{429 \ln2}{181}\right)\bar q_1 \bar q_2
+ \left(\frac{811 \ln2}{181}-\frac{541}{181}-2 \ln^22\right)\bar q_1 \bar \epsilon _1
+ \left(\frac{293}{181}+2\ln^22-\frac{315 \ln4}{181}\right)\bar \epsilon _1^2
\nb\\&&
~~
+ \left(-\frac{56}{181}-\ln^22+\frac{315 \ln2}{181}\right)\bar q_1 \bar \epsilon _2
+ \left(\frac{\pi ^2}{12}-\frac{4231}{1629}+\ln^22+\frac{47 \ln2}{181}\right)\bar \epsilon _1 \bar \epsilon _2
+ \left(-\frac{11}{362}+\frac{\ln^22}{2}-\frac{67 \ln2}{181}\right)\bar \epsilon _2^2
\nb\\&&~~
+ \left(\frac{\pi ^2}{24}-\frac{86}{1629}-\frac{\ln^22}{2} +\frac{67 \ln2}{181}\right)\bar \epsilon _2 \bar \epsilon _3+\mathcal{O}(\epsilon^3)\Bigg\}.
\eqn

Let us now compare the above expression with the one obtained by using the first-order uniform asymptotic approximation in Ref.\  \cite{Martin2013JCAP,Temme}, which reads
\bqn\lb{Martin-scalar}
\Delta_{s}^2(k)&\simeq&\frac{18 \bar H^2}{8\pi^2 e^3 M^2_{\text{pl}}\bar \epsilon_1 \bar c_0}\Bigg\{1+\left(\frac{7}{3}-\ln2\right)\bar q_1+\left(\ln4-\frac{8}{3}\right)\bar \epsilon_1+\left(\ln2-\frac{1}{3}\right)\bar \epsilon_2+\left(\frac{23}{18}-\frac{4\ln2}{3}+\frac{\ln^22}{2}\right)\bar q_1^2\nb\\
&&~~~~~+\left(\frac{25}{9}-\frac{\pi^2}{24}-\frac{7\ln2}{3}+\frac{\ln^22}{2}\right)\bar q_1\bar q_2+\left(\frac{13\ln2}{3}-2\ln^22-\frac{25}{9}\right)\bar q_1 \bar \epsilon_1+\left(\frac{13}{9}-\frac{10\ln2}{3}+2\ln^22\right)\bar \epsilon_1^2\nb\\
&&~~~~~+\left(\frac{5\ln2}{3}-\ln^22-\frac{2}{9}\right)\bar \epsilon_2\bar q_1+\left(\frac{\pi^2}{12}-\frac{25}{9}+\frac{\ln2}{3}+\ln^22\right)\bar \epsilon_1\bar \epsilon_2\nb\\
&&~~~~~+\left(\frac{\ln^22}{2}-\frac{\ln2}{3}-\frac{1}{18}\right)\bar \epsilon_2^2+\left(\frac{\pi^2}{24}-\frac{1}{9}+\frac{\ln2}{3}-\frac{\ln^22}{2}\right)\bar \epsilon_2\bar \epsilon_3+\mathcal{O}(\epsilon^3)\Bigg\}.
\eqn
In the overall amplitude, we notice that the two results have about  $10\%$ relative errors. The comparison of the numerical coefficients in the front of the Hubble flow parameters is presented in Table I, from which one can see clearly that the results of the first-order uniform asymptotic approximation have been significantly improved by our expression with high-order corrections from the high-order asymptotic approximations.

\begin{table}[htdp]
\caption{Comparison of the results obtained by using the first-order uniform asymptotic approximation \cite{Martin2013JCAP,Temme} with those obtained in this paper by using the third-order uniform asymptotic approximation.}
\begin{center}
\begin{tabular}{|c|c|c|c|c|c|c|c|c|c|c|c|}
\hline
Methods & $\bar q_1$ & $\bar \epsilon_1$ & $\bar \epsilon_2 $ & $\bar q_1^2$ &$\bar q_1 \bar q_2$ & $\bar q_1 \bar \epsilon_1$ & $\bar \epsilon_1^2$ & $\bar q_1 \bar \epsilon_2$ & $\bar \epsilon_1 \bar \epsilon_2 $ & $\bar \epsilon_2^2$ & $\bar \epsilon_2 \bar \epsilon_3$\\
\hline
1st-order & 1.64019 &  -1.28037 & 0.359814 & 0.59381& 1.17261& -0.844097& 0.094860 & 0.45257 & -1.24381 & -0.0463781 & 0.290945\\
\hline
3rd-order &1.67702 & -1.32089 & 0.322981 & 1.18458 &0.989427 & -0.735046 & 0.167079 & 0.416461 & -1.11439 & -0.0467396 &0.374793\\
\hline
Relative difference  & 5.7\%& 3.1\%& 10\% &  50\% &16\% & 13\%& 43\% & 8.7\% & 12\% &0.8\% & 22\%\\
\hline
\end{tabular}
\end{center}
\label{default}
\end{table}%

Now we turn to the scalar spectral index $n_s$. As the scalar spectrum is calculated up to the second-order in the uniform asymptotic approximation, the corresponding scalar spectral index can be calculated up to the third-order in the slow-roll approximation, which can be rewritten in the form,
\bqn
n_s-1&\simeq&
\bar q_1
-2 \bar \epsilon _1
-\bar \epsilon _2
-\bar q_1^2
+ \left(\frac{64}{27}-\ln2\right)\bar q_1 \bar q_2
+3 \bar q_1 \bar \epsilon _1
-2\bar  \epsilon _1^2
+\bar q_1 \bar \epsilon _2
+ \left(\ln4-\frac{101}{27}\right)\bar \epsilon _1 \bar \epsilon _2
+ \left(\ln2-\frac{10}{27}\right)\bar \epsilon _2 \bar \epsilon _3\nb\\
&&
+\bar q_1^3
+\left(\frac{913 \ln2}{315}-\frac{388}{81} +\ln8\right)\bar q_1^2 \bar q_2
+ \left(\frac{260}{81}-\frac{\pi ^2}{24}+\frac{\ln^22}{2}-\frac{64 \ln2}{27}\right)\bar q_1 \bar q_2^2
-4 \bar q_1^2 \bar \epsilon _1
+5 \bar q_1 \bar \epsilon _1^2
-2 \bar \epsilon _1^3
-\bar q_1^2\bar \epsilon _2
\nb\\&&
+\left(\frac{260}{81}-\frac{\pi ^2}{24}+\frac{\ln^22}{2}-\frac{64 \ln2}{27}\right)\bar q_1 \bar q_2 \bar q_3
+ \left(\frac{584}{81}-4 \ln2\right)\bar q_1 \bar q_2 \bar \epsilon _1
+\frac{11}{81} \bar \epsilon _2^2 \bar \epsilon _3
+ \left(\frac{100}{81}-\ln2\right)\bar q_1 \bar q_2 \bar \epsilon _2
\nb\\&&
+\left(\frac{695}{81}-5 \ln2\right)\bar q_1 \bar \epsilon _1 \bar \epsilon _2
 + \left(6\ln2-\frac{703}{81}\right)\bar \epsilon _1^2 \bar\epsilon _2
 +\left(\frac{49}{81}-\ln4\right)\bar q_1 \bar \epsilon _2 \bar \epsilon _3
  + \left(\frac{\pi ^2}{12}-\frac{167}{54}-\ln^22+\frac{128 \ln2}{27}\right)\bar\epsilon _1 \bar \epsilon _2 \bar \epsilon _3
\nb\\&&
+ \left(\frac{\pi ^2}{24}+\frac{19}{324}-\frac{ \ln^22}{2}+\frac{5\ln4)}{27}\right)\bar \epsilon _2 \bar \epsilon _3^2+ \left(\frac{\pi ^2}{24}+\frac{19}{324}-\frac{\ln^22}{2} +\frac{10 \ln2}{27}\right)\bar \epsilon _2 \bar \epsilon _3 \bar \epsilon _4+\mathcal{O}(\epsilon^4).
\eqn

Similarly, the running of the scalar spectral index up to the fourth-order in the slow-roll approximation, is given by
\bqn
\alpha_s&\simeq&
\bar q_1 \bar q_2
-2\bar \epsilon _1 \bar \epsilon _2
- \bar \epsilon _2 \bar \epsilon _3
+4 \bar q_2 \bar \epsilon _1 \bar q_1
+\bar q_2 \bar \epsilon_2 \bar q_1
-3 \bar q_2 \bar q_1^2
-6 \bar \epsilon_1^2 \bar \epsilon_2
+2 \bar \epsilon_2 \bar \epsilon_3 \bar q_1
+5 \bar \epsilon_1 \bar \epsilon_2 \bar q_1
+\left(\frac{64}{27}-\ln2\right) \bar q_2 \bar q_3 \bar q_1
\nb\\
&&
+\left(\ln4-\frac{101}{27}\right) \bar \epsilon_1 \bar \epsilon_2^2
+\left(\frac{64}{27}-\ln2\right) \bar q_1\bar q_2^2
+\left(\ln2-\frac{10}{27}\right) \bar \epsilon_2 \bar \epsilon_3^2
+\left(\ln4-\frac{128}{27}\right) \bar \epsilon_1 \bar \epsilon_2 \bar \epsilon_3
+\left(\ln2-\frac{10}{27}\right) \bar \epsilon_2 \bar \epsilon_3 \bar \epsilon_4
\nb\\
&&
+6 \bar q_2 \bar q_1^3
-3 \bar q_2 \bar \epsilon_2 \bar q_1^2
-9 \bar \epsilon_1 \bar \epsilon_2 \bar q_1^2
+\left(\frac{2141 \ln2}{315}-\frac{1022}{81}\right) \bar q_2^2 \bar q_1^2
+\left(\frac{1228 \ln2}{315}-\frac{580}{81}\right) \bar q_2 \bar q_3 \bar q_1^2
-15 \bar q_2 \bar \epsilon_1 \bar q_1^2
-3 \bar \epsilon_2 \bar \epsilon_3 \bar q_1^2
%
\nb\\
&&
%
+\left(\frac{\ln^22}{2}-\frac{64 \ln8}{81}-\frac{\pi ^2}{24}+\frac{260}{81}\right) \bar q_2^3 \bar q_1
+\left(\frac{\ln^22}{2}-\frac{64 \ln8}{81}-\frac{\pi ^2}{24}+\frac{260}{81}\right) \bar q_2 \bar q_3^2 \bar q_1
+\left(\frac{998}{81}-7 \ln2\right) \bar \epsilon_1 \bar \epsilon_2^2 \bar q_1
\nb\\
&&
+9 \bar q_2 \bar \epsilon_1^2 \bar q_1
+\left(\frac{3 \ln^22}{2}-\frac{64 \ln2}{9}-\frac{\pi ^2}{8}+\frac{260}{27}\right) \bar q_2^2 \bar q_3 \bar q_1
+\left(\frac{776}{81}-5 \ln2\right) \bar q_2 \bar q_3 \bar \epsilon_1 \bar q_1
+\left(\frac{1495}{81}-9 \ln2\right) \bar q_2 \bar \epsilon_1 \bar \epsilon_2 \bar q_1
%
\nb\\&&
+\left(\frac{\ln^22}{2}-\frac{64 \ln8}{81}-\frac{\pi ^2}{24}+\frac{260}{81}\right) \bar q_2 \bar q_3 \bar q_4 \bar q_1
+ \left(\frac{776}{81}-5  \ln2\right)\bar \epsilon_1 \bar q_1 \bar q_2^2
+\left(\frac{100}{81}-\ln2\right) \bar q_2^2 \bar \epsilon_2 \bar q_1
+21 \bar \epsilon_1^2 \bar \epsilon_2 \bar q_1
\nb\\
&&
%
+\left(\frac{100}{81}-\ln2\right) \bar q_2 \bar q_3 \bar \epsilon_2 \bar q_1
+\left(\frac{176}{81}-\ln8\right) \bar q_2 \bar \epsilon_2 \bar \epsilon_3 \bar q_1
+\left(\frac{1241}{81}-7 \ln2\right) \bar \epsilon_1 \bar \epsilon_2 \bar \epsilon_3 \bar q_1+\left(14 \ln2-\frac{1763}{81}\right) \bar \epsilon_1^2 \bar \epsilon_2^2
%
\nb\\
&&
+\left(\frac{79}{81}-\ln8\right) \bar \epsilon_2 \bar \epsilon_3 \bar \epsilon_4 \bar q_1
+\left(\frac{101 \ln2}{27}-\ln^22+\frac{\pi ^2}{12}-\frac{67}{18}\right) \bar \epsilon_1 \bar \epsilon_2^3
+\left(\frac{10 \ln2}{27}-\frac{ \ln^22}{2}+\frac{\pi ^2}{24}+\frac{19}{324}\right) \bar \epsilon_2 \bar \epsilon_3^3
%
\nb\\
&&
%
+\left(\frac{155 \ln2}{27}-\ln^22+\frac{\pi ^2}{12}-\frac{187}{54}\right) \bar \epsilon_1 \bar \epsilon_2 \bar \epsilon_3^2
+\frac{49}{81} \bar \epsilon_2^2 \bar \epsilon_3^2
+\left(\frac{10 \ln2}{27}-\frac{\ln^22}{2} +\frac{\pi ^2}{24}+\frac{19}{324}\right) \bar \epsilon_2 \bar \epsilon_3 \bar \epsilon_4^2
-12 \bar \epsilon_1^3 \bar \epsilon_2
%
\nb\\
&&
%
%
+\left(\frac{110 \ln2}{9}-3 \ln^22+\frac{\pi ^2}{4}-\frac{551}{54}\right) \bar \epsilon_1 \bar \epsilon_2^2 \bar \epsilon_3
+\frac{11}{81} \bar \epsilon_2^2 \bar \epsilon_3 \bar \epsilon_4
+\left(8\ln2-\frac{1087}{81}\right) \bar \epsilon_1^2 \bar \epsilon_2 \bar \epsilon_3
+\left(\frac{79}{81}-\ln8\right) \bar \epsilon_2 \bar \epsilon_3^2 \bar q_1
\nb\\
&&
%
+\left(\frac{10 \ln2}{9}-\frac{3 \ln^22}{2} +\frac{\pi ^2}{8}+\frac{19}{108}\right) \bar \epsilon_2 \bar \epsilon_3^2 \bar \epsilon_4
+\left(\frac{155 \ln2}{27}-\ln^22+\frac{\pi ^2}{12}-\frac{187}{54}\right) \bar \epsilon_1 \bar \epsilon_2 \bar \epsilon_3 \bar \epsilon_4
\nb\\
&&
%
+\left(\frac{10 \ln2}{27}-\frac{\ln^22}{2} +\frac{\pi ^2}{24}+\frac{19}{324}\right) \bar \epsilon_2 \bar \epsilon_3 \bar \epsilon_4 \bar \epsilon_5+\mathcal{O}(\epsilon^5).
\eqn

\subsection{Tensor Perturbations}

For the tensor perturbations, we can simply repeat the above analysis to calculate its spectrum, spectral index, and running of the spectral index. However, this has actually been already done in \cite{ZhuUniform2014-2}, because the equation of motion for the tensor perturbations in the k-inflation is the same as that in general relativity calculated explicitly in \cite{ZhuUniform2014-2}. However, those expressions were written only in terms of the the slow-roll parameters $(\epsilon,\;\delta_1,\;\delta_2,\;\cdots)$. Therefore, in the following, we only need to write those expressions in terms of the slow-roll parameters $(\epsilon_1,\;\epsilon_2,\;\cdots)$.
In particular, we find that the  tensor spectrum take the form,
\bqn
\Delta_t^2(k) &\simeq& \frac{181 \bar H^2}{36 e^3 \pi^2} \Bigg\{1+ \left(\ln4-\frac{496}{181}\right)\bar \epsilon _1
+ \left(\frac{293}{181}+2 \ln^22-\frac{630\ln2}{181}\right)\bar \epsilon _1^2\nb\\
&&~~~~~~~~~~~~
+\left(\frac{\pi ^2}{12}-\frac{4636}{1629}-\ln^22+\frac{496 \ln2}{181}\right) \bar \epsilon _1\bar \epsilon _2+\mathcal{O}(\epsilon^3) \Bigg\}.
\eqn
Similar to the case for the scalar spectrum, here we compare our results with those obtained by using the first-order uniform approximation \cite{Martin2013JCAP,Temme},
\bqn
\mathcal{P}_{h,\text{1st-uniform}}(k) &\simeq& \frac{36\bar H^2}{e^3 \pi^2 M_{\text{pl} }^2 } \Bigg\{1+\left(2\ln2-\frac{8}{3}\right)\bar \epsilon_1+\left(\frac{13}{9}-\frac{10\ln2}{3}+2\ln^22\right)\bar \epsilon^2_1\nb\\
&&~~~~~~~~~~~~~~~+\left(\frac{\pi^2}{12}-\frac{26}{9}+\frac{8\ln2}{3}-\ln^22\right)\bar\epsilon_1 \bar \epsilon_2 \Bigg\}.
\eqn
By identifying $ \mathcal{P}_{h,\text{1st-uniform}}\sim 8 \Delta_t^2(k)$ and comparing the amplitudes of the spectra, we find that the results by the first-order uniform asymptotic approximation have about $10\%$ relative errors. The comparison of the numerical coefficients in the front of the Hubble flow parameters are listed in Table II. Similar to the case for the scalar spectrum, the high-order corrections of the third-order uniform asymptotic approximation make significant improvement over the results obtained from the first-order uniform asymptotic approximation.

\begin{table}[htdp]
\caption{Comparing the results obtained by using the first-order uniform approximation \cite{Martin2013JCAP} with those obtained in this paper by using the third-order uniform asymptotic approximation}
\begin{center}
\begin{tabular}{|c|c|c|c|}
\hline
Methods &  $\bar \epsilon_1$ & $\bar \epsilon_1^2$  & $\bar \epsilon_1 \bar \epsilon_2 $ \\
\hline
1st-order &   -1.28037 & 0.094860 & -1.24381 \\
\hline
3rd-order  & -1.32089  & 0.167079  & -0.604451 \\
\hline
Relative difference  & 3.1\%& 43\%  & 13\%  \\
\hline
\end{tabular}
\end{center}
\label{default}
\end{table}%

The tensor spectral index, on the other hand, is given by
\bqn
n_t&\simeq&
-2 \bar \epsilon_1
-2 \bar \epsilon_1^2
+\left(2 \ln2-\frac{74}{27}\right)\bar \epsilon_1\epsilon_2
-2 \bar \epsilon_1^3
+ \left(\frac{\pi ^2}{12}-\frac{425}{162}-\ln^22+\frac{74 \ln2}{27}\right) \bar \epsilon_1\bar \epsilon_2^2\nb\\
&&
+ \left(\frac{\pi ^2}{12}-\frac{425}{162}-\ln^22+\frac{74 \ln2}{27}\right)\bar \epsilon_1\bar \epsilon_2  \bar \epsilon_3 
+ \left(6 \ln2-\frac{622}{81}\right) \bar \epsilon_1^2\bar \epsilon_2+\mathcal{O}(\epsilon^4).
\eqn
The running of the spectral index reads
\bqn
\alpha_t &\simeq&
-2 \bar \epsilon_1 \bar \epsilon_2
-6 \bar \epsilon_1^2 \bar \epsilon_2
+ \left(2 \ln2-\frac{74}{27}\right)\bar \epsilon_1 \bar \epsilon_2^2
+ \left(2 \ln2-\frac{74}{27}\right)\bar \epsilon_1 \bar \epsilon_2 \bar \epsilon_3
-12 \bar \epsilon_1^3 \bar \epsilon_2
+ \left(14 \ln2-\frac{1520}{81}\right)\bar \epsilon_1^2 \bar \epsilon_2^2
\nb\\
&&
+ \left(\frac{\pi ^2}{12}-\frac{425}{162}-\ln^22+\frac{74 \ln2}{27}\right)\bar \epsilon_1 \bar \epsilon_2^3
+ \left(8 \ln2-\frac{844}{81}\right)\bar \epsilon_1^2 \bar \epsilon_2 \bar \epsilon_3
+ \left(\frac{\pi ^2}{4}-\frac{425}{54}-3 \ln^22+\frac{74 \ln2}{9}\right)\bar \epsilon_1 \bar \epsilon_2^2 \bar \epsilon_3\nb\\
&&
+ \left(\frac{\pi ^2}{12}-\frac{425}{162}-\ln^22+\frac{74 \ln2}{27}\right)\bar \epsilon_1 \bar \epsilon_2 \bar \epsilon_3^2
+\left(\frac{\pi ^2}{12}-\frac{425}{162}-\ln^22+\frac{74 \ln2}{27}\right)\bar \epsilon_1 \bar \epsilon_2 \bar \epsilon_3 \bar \epsilon_4 +\mathcal{O}(\epsilon^5).
\eqn

\subsection{Expressions in terms of quantities calculated at horizon crossing}

So far, we have obtained all the expressions of the power spectra, spectral indices, and running of spectral indices for both the scalar and tensor perturbations in the k-inflation, evaluated only at the turning point. However, usually those expressions were expressed in terms of the slow-roll parameters which are evaluated at the time $\eta_\star$  when scalar or tensor modes cross the horizon, i.e., $a(\eta_\star) H(\eta_\star)= c_s(\eta_\star) k$ for scalar perturbations and  $c^2_s(\eta_\star)=1$ for tensor perturbations. Consider modes with the same wavenumber $k$, it is easy to see that the scalar mode and tensor mode left horizon at different times if $c^2_s(\eta_\star)\neq 1$.  When $c^2_s(\eta_\star)>1$, the scalar mode leaves horizon later than the tensor mode, and for $c_s^2(\eta_\star) <1$, the scalar mode leaves horizon before the tensor one does. In this case,  caution must be taken for the evaluation time for all the inflationary observables. As we have two different horizon crossing times, it is reasonable to re-write all our results in terms of quantities evaluated at the later time, i.e., we should evaluate all expressions at scalar horizon crossing $a(\eta_\star) H(\eta_\star)= c_s(\eta_\star) k$ for $c_s^2(\eta_\star)>1$ and at tensor horizon crossing $a(\eta_\star) H(\eta_\star)= k$ for $c_s^2(\eta_\star)<1$. In the following, we present all the expressions for both cases, respectively.

\subsubsection{$c_s^2(\eta_\star)>1$}
For $c_s^2(\eta_\star)>1$, as the scalar mode leaves horizon later than the tensor mode, we shall re-write all the expressions in terms of quantities evaluated at the time when scalar leaves the Hubble horizon $a(\eta_\star) H(\eta_\star)=c_s(\eta_\star)k$. 
Skipping all the tedious calculations, we find that the scalar spectrum can be written in the form
\bqn\lb{scalarUniform}
\Delta^2_s(k) &\simeq& \frac{181H_\star ^2}{72 e^3\pi^2 \epsilon_{\star1} c_{\star0}} \Bigg\{1+ \left(\frac{429}{181}-\ln3\right)q_{\star1}+ \left(\ln9-\frac{496}{181}\right)\epsilon _{\star1}+ \left(\ln3-\frac{67}{181}\right)\epsilon _{\star2}\nb\\
&&~~~~~~~~~~~~~~~~
+\left(\frac{457}{362}+\frac{\ln^23}{2}-\frac{248 \ln3}{181}-\frac{64 \ln2}{1267}\right)q_{\star1}^2
+ \left(\frac{517}{543}+2 \ln^23-\frac{630 \ln3}{181}\right)\epsilon _{\star1}^2\nb\\
&&~~~~~~~~~~~~~~~~
+ \left(\frac{\pi ^2}{12}-\frac{3688}{1629}+\ln^23+\frac{47 \ln3}{181}\right)\epsilon _{\star1} \epsilon _{\star2}+ \left(\frac{329}{1086}+\frac{\ln^23}{2}-\frac{67 \ln3}{181}\right)\epsilon _{\star2}^2\nb\\
&&~~~~~~~~~~~~~~~~
+ \left(-\frac{\pi ^2}{24}+\frac{4865}{1629}+\frac{\ln^23}{2}-\frac{429 \ln3}{181}\right)q_{\star1}q_{\star2} + \left(-\frac{718}{543}-2 \ln^23+\frac{811 \ln3}{181}\right)q_{\star1}\epsilon _{\star1}\nb\\
&&~~~~~~~~~~~~~~~~
+ \left(\frac{13}{543}-\ln^23+\frac{315 \ln3}{181}\right)q_{\star1} \epsilon _{\star2}+ \left(\frac{\pi ^2}{24}-\frac{86}{1629}-\frac{ \ln^23}{2}+\frac{67 \ln3}{181}\right)\epsilon _{\star2} \epsilon _{\star3}+\mathcal{O}(\epsilon^3)\Bigg\}.\nb\\
\eqn
Note that in the above the subscript $\star$ represents the quantities evaluated at $\eta_\star$. Now we turn to the scalar spectral index, which can be written as
\bqn
n_s-1&\simeq& q_{\star1}-2 \epsilon _{\star1}-\epsilon _{\star2}+3 q_{\star1}\epsilon _{\star1}+q_{\star1} \epsilon _{\star2}-q_{\star1}^2-2 \epsilon _{\star1}^2+\left(\frac{64}{27}-\ln3\right)q_{\star1} q_{\star2}+ \left(\ln9-\frac{101}{27}\right)\epsilon_{\star 1} \epsilon _{\star2}\nb\\
&&
+\left(\ln3-\frac{10}{27}\right)\epsilon _{\star2} \epsilon _{\star3} 
-4 q_{\star1}^2 \epsilon _{\star1}
-q_{\star1}^2 \epsilon _{\star2}
+5 q_{\star1} \epsilon _{\star1}^2
+ \left(\frac{611}{81}-4 \ln3\right) q_{\star1}q_{\star2}  \epsilon _{\star1}
+ \left(\frac{73}{81}-\ln3\right) q_{\star1}q_{\star2}  \epsilon _{\star2}\nb\\
&&
+ \left(\frac{803}{81}-5 \ln3\right)q_{\star1} \epsilon _{\star1} \epsilon _{\star2}
+\left(\frac{103}{81}-2 \ln3\right)q_{\star1} \epsilon _{\star2} \epsilon _{\star3} 
+q_{\star1}^3
+\left(\frac{260}{81}-\frac{\pi ^2}{24}+\frac{\ln^23}{2}-\frac{64 \ln3}{27}\right)q_{\star1}q_{\star2}^2  \nb\\
&&
+ \left(-\frac{\pi ^2}{24}+\frac{260}{81}+\frac{\ln^23}{2}-\frac{64 \ln3}{27}\right)q_{\star1}q_{\star2} q_{\star3} 
+ \left(3 \ln3-\frac{442}{81}-\frac{32 \ln2}{315}\right)q_{\star1}^2q_{\star2} 
-2 \epsilon _{\star1}^3+\frac{38}{81} \epsilon _{\star2}^2 \epsilon _{\star3}\nb\\
&&
+\left(\frac{\pi ^2}{12}-\frac{55}{18}-\ln^23+\frac{101 \ln3}{27}\right)\epsilon _{\star1} \epsilon _{\star2}^2 
+ \left(\frac{\pi ^2}{24}+\frac{19}{324}-\frac{\ln^23}{2} +\frac{10 \ln3}{27}\right)\epsilon _{\star2} \epsilon _{\star3}^2
+ \left(6 \ln3-\frac{757}{81}\right)\epsilon _{\star1}^2 \epsilon _{\star2}\nb\\
&&
+ \left(\frac{\pi ^2}{12}-\frac{185}{54}-\ln^23+\frac{128 \ln3}{27}\right)\epsilon _{\star1} \epsilon _{\star2} \epsilon _{\star3}+\left(\frac{\pi ^2}{24}+\frac{19}{324}-\frac{\ln^23}{2} +\frac{10 \ln3}{27}\right)\epsilon _{\star2} \epsilon _{\star3} \epsilon _{\star4}+\mathcal{O}(\epsilon^4).
\eqn

Finally,  the running of the scalar spectral index can be determined up to the fourth-order in terms of  the slow-roll parameters, and is given by
\bqn
\alpha_s&\simeq&
q_{\star2} q_{\star1}-\epsilon _{\star2} \epsilon _{\star3}-2 \epsilon _{\star1} \epsilon _{\star2}
+4 q_{\star2} \epsilon _{\star1} q_{\star1}
+5 \epsilon _{\star1} \epsilon _{\star2} q_{\star1}
+q_{\star2} \epsilon _{\star2} q_{\star1}
+2 \epsilon _{\star2} \epsilon _{\star3} q_{\star1}
-6 \epsilon _{\star1}^2 \epsilon _{\star2}
-3 q_{\star2} q_{\star1}^2\nb\\
&&
+\left(\frac{64}{27}-\ln3\right) q_{\star2}^2 q_{\star1}
+\left(\frac{64}{27}-\ln3\right) q_{\star2} q_{\star3} q_{\star1}
+\left(\ln9-\frac{101}{27}\right) \epsilon _{\star1} \epsilon _{\star2}^2
+\left(\ln3-\frac{10}{27}\right) \epsilon _{\star2} \epsilon _{\star3}^2\nb\\
&&
+\left(\ln9-\frac{128}{27}\right) \epsilon _{\star1} \epsilon _{\star2} \epsilon _{\star3}
+\left(\ln3-\frac{10}{27}\right) \epsilon _{\star2} \epsilon _{\star3} \epsilon _{\star4}
+6 q_{\star2} q_{\star1}^3
+\left(7 \ln3-\frac{64 \ln2}{315}-\frac{1076}{81}\right) q_{\star2}^2 q_{\star1}^2\nb\\
&&
+\left(2\ln9-\frac{634}{81}-\frac{32 \ln2}{315} \right) q_{\star2} q_{\star3} q_{\star1}^2
-15 q_{\star2} \epsilon _{\star1} q_{\star1}^2
-3 q_{\star2} \epsilon _{\star2} q_{\star1}^2
-9 \epsilon _{\star1} \epsilon _{\star2} q_{\star1}^2
-3 \epsilon _{\star2} \epsilon _{\star3} q_{\star1}^2\nb\\
&&
+\left(\frac{\ln^23}{2}-\frac{64 \ln3}{27}-\frac{\pi ^2}{24}+\frac{260}{81}\right) q_{\star2}^3 q_{\star1}
+\left(\frac{\ln^23}{2}-\frac{64 \ln3}{27}-\frac{\pi ^2}{24}+\frac{260}{81}\right) q_{\star2} q_{\star3}^2 q_{\star1}
+9 q_{\star2} \epsilon _{\star1}^2 q_{\star1}\nb\\
&&
+\left(\frac{1106}{81}-7 \ln3\right) \epsilon _{\star1} \epsilon _{\star2}^2 q_{\star1}
+\left(\frac{3 \ln^23}{2}-\frac{64 \ln3}{9}-\frac{\pi ^2}{8}+\frac{260}{27}\right) q_{\star2}^2 q_{\star3} q_{\star1}
+\left(\frac{133}{81}-3 \ln3\right) \epsilon _{\star2} \epsilon _{\star3}^2 q_{\star1}
\nb\\
&&
%
+\left(\frac{1}{2} \ln^2\frac{3}{2}+\frac{\ln^22}{2}-\frac{\pi ^2}{24}-\frac{64 \ln8}{81}+\frac{260}{81}\right) q_{\star2} q_{\star3} q_{\star4} q_{\star1}
+\left(\frac{803}{81}-5 \ln3\right) q_{\star2}^2 \epsilon _{\star1} q_{\star1}
+21 \epsilon _{\star1}^2 \epsilon _{\star2} q_{\star1}
\nb\\&&
%
+\left(\frac{73}{81}-\ln3\right) q_{\star2}^2 \epsilon _{\star2} q_{\star1}
+\left(\frac{803}{81}-5 \ln3\right) q_{\star2} q_{\star3} \epsilon _{\star1} q_{\star1}
+\left(\frac{73}{81}-\ln3\right) q_{\star2} q_{\star3} \epsilon _{\star2} q_{\star1}
-12 \epsilon _{\star1}^3 \epsilon _{\star2}\nb\\
&&
+\left(\frac{176}{81}-3 \ln3\right) q_{\star2} \epsilon _{\star2} \epsilon _{\star3} q_{\star1}
+\left(\frac{1349}{81}-7 \ln3\right) \epsilon _{\star1} \epsilon _{\star2} \epsilon _{\star3} q_{\star1}
+\left(\frac{101 \ln3}{27}-\ln^23+\frac{\pi ^2}{12}-\frac{55}{18}\right) \epsilon _{\star1} \epsilon _{\star2}^3
\nb\\&&
%
+\left(\frac{133}{81}-3 \ln3\right) \epsilon _{\star2} \epsilon _{\star3} \epsilon _{\star4} q_{\star1}
+\left(\frac{10 \ln3}{27}-\frac{\ln^23}{2} +\frac{\pi ^2}{24}+\frac{19}{324}\right) \epsilon _{\star2} \epsilon _{\star3}^3
+\left(14 \ln3-\frac{1817}{81}\right) \epsilon _{\star1}^2 \epsilon _{\star2}^2
%
\nb\\&&
+\frac{76}{81} \epsilon _{\star2}^2 \epsilon _{\star3}^2
+\left(\frac{1495}{81}-9 \ln3\right) q_{\star2} \epsilon _{\star1} \epsilon _{\star2} q_{\star1}
+\left(\frac{155 \ln3}{27}-\ln^23+\frac{\pi ^2}{12}-\frac{205}{54}\right) \epsilon _{\star1} \epsilon _{\star2} \epsilon _{\star3}^2
\nb\\&&
+\left(\frac{110 \ln3}{9}-3 \ln^23++\frac{\pi ^2}{4}-\frac{515}{54}\right) \epsilon _{\star1} \epsilon _{\star2}^2 \epsilon _{\star3}
+\left(8 \ln3-\frac{1141}{81}\right) \epsilon _{\star1}^2 \epsilon _{\star2} \epsilon _{\star3}
+\frac{38}{81} \epsilon _{\star2}^2 \epsilon _{\star3} \epsilon _{\star4}
\nb\\&&
%
+\left(\frac{10 \ln3}{9}-\frac{3 \ln^23}{2} +\frac{\pi ^2}{8}+\frac{19}{108}\right) \epsilon _{\star2} \epsilon _{\star3}^2 \epsilon _{\star4}
+\left(\frac{155 \ln3}{27}-\ln^23+\frac{\pi ^2}{12}-\frac{205}{54}\right) \epsilon _{\star1} \epsilon _{\star2} \epsilon _{\star3} \epsilon _{\star4}
\nb\\&&
+\left(\frac{10 \ln3}{27}-\frac{\ln^23}{2} +\frac{\pi ^2}{24}+\frac{19}{324}\right) \epsilon _{\star2} \epsilon _{\star3} \epsilon _{\star4} \epsilon _{\star5}
+\left(\frac{10 \ln3}{27}-\frac{\ln^23}{2} +\frac{\pi ^2}{24}+\frac{19}{324}\right) \epsilon _{\star2} \epsilon _{\star3} \epsilon _{\star4}^2+\mathcal{O}(\epsilon^5).\nb\\
\eqn

Now let us turn to consider the tensor perturbations. For the tensor spectrum, we get
\bqn
\Delta_t^2(k) &\simeq& \frac{181 H_\star^2}{36e^3 \pi^2} \Bigg\{1+\left(\ln(9c_{\star0})-\frac{496}{181}\right)\epsilon _{\star1} +\left(\frac{517}{543}+2 \ln^2(3c_{\star0})-\frac{630\ln(3c_{\star0})}{181}\right)\epsilon _{\star1}^2 \nb\\
&&~~~~~~~~~~~~+ \left(\frac{\pi ^2}{12}-\frac{4636}{1629}-\ln^2(3c_{\star0})+\frac{496 \ln(3c_{\star0})}{181}\right)\epsilon _{\star2} \epsilon _{\star1}+\mathcal{O}(\epsilon^3)\Bigg\}.\nb\\
\eqn

The tensor spectral index $n_t$, on the other hand, can be  expressed as
\bqn
n_t &\simeq&
-2 \epsilon _{\star1}
-2\epsilon _{\star1}^2
+\left(2 \ln(3c_{\star0})-\frac{74}{27}\right)\epsilon_{\star1}\epsilon_{\star2}
-2 \epsilon _{\star1}^3
+  \left(\frac{\pi ^2}{12}-\frac{425}{162}-\ln^2(3c_{\star0})+\frac{74 \ln(3c_{\star0})}{27}\right)\epsilon _{\star1}\epsilon _{\star2}^2
\nb\\
&&
+ \left(6 \ln(3c_{\star0})-\frac{676}{81}\right) \epsilon _{\star1}^2\epsilon _{\star2}
+ \left(\frac{\pi ^2 }{12}-\frac{425 }{162}- \ln^2(3c_{\star0})+\frac{74 \ln(3c_{\star0})}{27} \right)\epsilon _{\star1}\epsilon _{\star2}\epsilon _{\star3}+\mathcal{O}(\epsilon^4),
\eqn
while  the running of the tensor spectral index is given by
\bqn
\alpha_t &\simeq &
-2  \epsilon _{\star1}\epsilon _{\star2}
-6 \epsilon _{\star2} \epsilon _{\star1}^2
+ \left(2 \ln(3c_{\star0})-\frac{74}{27}\right) \epsilon _{\star1}\epsilon _{\star2}^2
+\left(2 \ln(3c_{\star0})-\frac{74}{27}\right) \epsilon _{\star1} \epsilon _{\star2}\epsilon _{\star3}
-12 \epsilon _{\star2} \epsilon _{\star1}^3
\nb\\&&
+ \left(\frac{\pi ^2}{12}-\frac{425}{162}-\ln^2(3c_{\star0})+\frac{74 \ln(3c_{\star0})}{27}\right)\epsilon _{\star2}^3 \epsilon _{\star1}
+ \left(8 \ln(3c_{\star0})-\frac{898}{81}\right)\epsilon _{\star2} \epsilon _{\star3}\epsilon _{\star1}^2
+ \left(14 \ln(3c_{\star0})-\frac{1574}{81}\right)\epsilon _{\star2}^2 \epsilon _{\star1}^2\nb\\
&&
+ \left(\frac{\pi ^2}{12}-\frac{425}{162}-\ln^2(3c_{\star0})+\frac{74 \ln(3c_{\star0})}{27}\right)\epsilon _{\star2} \epsilon _{\star3}^2 \epsilon _{\star1}
+\left(\frac{\pi ^2}{4}-\frac{425}{54}-3 \ln^2(3c_{\star0})+\frac{74 \ln(3c_{\star0})}{9}\right)\epsilon _{\star2}^2\epsilon _{\star3} \epsilon _{\star1} \nb\\
&&
+ \left(\frac{\pi ^2}{12}-\frac{425}{162}-\ln^23+\frac{74 \ln(3c_{\star0})}{27}\right)\epsilon _{\star2} \epsilon _{\star3}\epsilon _{\star4} \epsilon _{\star1}+\mathcal{O}(\epsilon^5).
\eqn

Finally, with both the scalar and tensor spectra given above, we can evaluate the tensor-to-scalar ratio at the time when the scalar mode leaves horizon, i.e., $a(\eta_\star) H_\star=c_s(\eta_\star) k$, and find that
\bqn
r&\simeq&
16 c_{\star0} \epsilon_{\star1} \Bigg\{1
+\left(\ln3-\frac{429}{181}\right)q_{\star1}+2\epsilon _{\star1} \ln c_{\star0}
+\left(\frac{67}{181}-\ln3\right)\epsilon _{\star2}
+ \left(2 \ln^2c_{\star0}+2 \ln c_{\star0}\right)\epsilon _{\star1}^2
\nb\\&&~~~~~~~~~~~
+ \left(\ln9 \ln c_{\star0}-\frac{858 \ln c_{\star0}}{181}-\frac{508394}{98283}+3\ln3\right)q_{\star1}\epsilon _{\star1}
+ \left(\frac{\pi ^2}{24}-\frac{4865}{1629}-\frac{\ln^23}{2} +\frac{429 \ln3}{181}\right)q_{\star1}q_{\star2}
\nb\\&&~~~~~~~~~~~
+\left(\frac{677 \ln3}{181}-\frac{174811}{98283}-\ln^23\right)q_{\star1}\epsilon _{\star2} 
+\left(\frac{630 \ln c_{\star0}}{181}+\frac{42500}{98283}-\ln3-\ln^2c_{\star0}-4 \ln3 \ln c_{\star0}\right)\epsilon _{\star1} \epsilon _{\star2} 
\nb\\&&~~~~~~~~~~~
+ \left(\frac{285365}{65522}+\frac{\ln^23}{2}-\frac{610 \ln3}{181}+\frac{64 \ln2}{1267}\right)q_{\star1}^2
+ \left(\frac{\ln^23}{2}-\frac{67 \ln3}{181}-\frac{32615}{196566}\right)\epsilon _{\star2}^2
\nb\\&&~~~~~~~~~~~
+ \left(-\frac{\pi ^2}{24}+\frac{86}{1629}+\frac{\ln^23}{2}-\frac{67 \ln3}{181}\right)\epsilon _{\star2}\epsilon _{\star3}+\mathcal{O}(\epsilon^3)\Bigg\}.
\eqn

\subsubsection{$c_s^2(\eta_\star)<1$}

For $c_s^2(\eta_\star)<1$, as the scalar mode leaves horizon before the tensor mode does, we shall re-write all the expressions in terms of quantities evaluated at the time when the tensor mode leaves the Hubble horizon $a(\eta_\star) H(\eta_\star)=k$. Skipping all the tedious calculations, we find that the scalar spectrum can be written in the form
\bqn\lb{scalarUniform}
\Delta^2_s(k) &\simeq&\frac{181 H_{\star }^2}{72 e^3 \pi ^2 c_{\star 0} \epsilon_{\star 1}} \Bigg\{ 1+ \left(\frac{429}{181}-\ln\frac{3}{c_{\star 0}}\right) q_{\star 1}+\left(-\frac{496}{181}+2 \ln\frac{3}{c_{\star 0}}\right) \epsilon_{\star 1}+\left(-\frac{67}{181}+\ln\frac{3}{c_{\star 0}}\right) \epsilon_{\star 2}\nb\\
&&~~+\left(\frac{457}{362}-\frac{64 \ln2}{1267}-\frac{248}{181} \ln\frac{3}{c_{\star 0}}+\frac{1}{2} \ln^2\frac{3}{c_{\star 0}}\right) q_{\star 1}^2+\left(\frac{4865}{1629}-\frac{\pi^2}{24}-\frac{429}{181} \ln\frac{3}{c_{\star 0}}+\frac{1}{2} \ln^2\frac{3}{c_{\star 0}}\right) q_{\star 1} q_{\star 2}\nb\\
&&~~+\left(-\frac{718}{543}+\frac{811}{181} \ln\frac{3}{c_{\star 0}}-2 \ln^2\frac{3}{c_{\star 0}}\right) q_{\star 1}\epsilon_{\star 1}
         +\left(\frac{517}{543}-\frac{630}{181} \ln\frac{3}{c_{\star 0}}+2 \ln^2\frac{3}{c_{\star 0}}\right) \epsilon_{\star 1}^2\nb\\
&&~~+\left(\frac{13}{543}+\frac{315}{181} \ln\frac{3}{c_{\star 0}}-\ln^2\frac{3}{c_{\star 0}}\right) q_{\star 1}\epsilon_{\star 2}
        +\left(-\frac{3688}{1629}+\frac{\pi ^2}{12}+\frac{47}{181} \ln\frac{3}{c_{\star 0}}+\ln^2\frac{3}{c_{\star 0}}\right) \epsilon_{\star 1} \epsilon _{\star 2}\nb\\
&&~~+\left(\frac{329}{1086}-\frac{67}{181} \ln\frac{3}{c_{\star 0}}+\frac{1}{2} \ln^2\frac{3}{c_{\star 0}}\right) \epsilon_{\star 2}^2
        +\left(-\frac{86}{1629}+\frac{\pi^2}{24}+\frac{67}{181} \ln\frac{3}{c_{\star 0}}-\frac{1}{2} \ln^2\frac{3}{c_{\star 0}}\right) \epsilon_{\star 2} \epsilon_{\star 3}+\mathcal{O}(\epsilon^3)\Bigg\}.\nb\\
\eqn
Now we turn to the scalar spectral index, which can be written as
\bqn
n_s-1&\simeq& q_{\star 1}-2 \epsilon_{\star 1}-\epsilon_{\star 2}-q_{\star 1}^2+\left(\frac{64}{27}-\ln\frac{3}{c_{\star 0}}\right) q_{\star 1} q_{\star 2}+3 q_{\star 1} \epsilon_{\star 1}-2 \epsilon_{\star 1}^2+q_{\star 1} \epsilon_{\star 2}+\left(-\frac{101}{27}+2 \ln\frac{3}{c_{\star 0}}\right) \epsilon_{\star 1} \epsilon_{\star 2}
\nb\\&&
+\left(-\frac{10}{27}+\ln\frac{3}{c_{\star 0}}\right) \epsilon_{\star 2} \epsilon_{\star 3}+q_{\star 1}^3-4 q_{\star 1}^2 \epsilon_{\star 1}+5 q_{\star 1} \epsilon_{\star 1}^2-2 \epsilon _{\star 1}^3-q_{\star 1}^2 \epsilon_{\star 2}+\frac{38}{81} \epsilon_{\star 2}^2 \epsilon _{\star 3}+\left(\frac{73}{81}-\ln\frac{3}{c_{\star 0}}\right) q_{\star 1}q_{\star 2} \epsilon_{\star 2}\nb\\
&&
+\left(-\frac{442}{81}-\frac{2867 \ln2}{315}+\frac{9 \ln4}{2}+3 \ln\frac{3}{c_{\star 0}}\right)q_{\star 1}^2 q_{\star 2}+\left(\frac{19}{324}+\frac{\pi ^2}{24}+\frac{10}{27} \ln\frac{3}{c_{\star 0}}-\frac{1}{2} \ln^2\frac{3}{c_{\star 0}}\right) \epsilon_{\star 2} \epsilon_{\star 3}^2\nb\\
&&
+\left(\frac{260}{81}-\frac{\pi ^2}{24}-\frac{64}{27} \ln\frac{3}{c_{\star 0}}+\frac{1}{2}\ln^2\frac{3}{c_{\star 0}}\right) q_{\star 1} q_{\star 2}^2+\left(\frac{260}{81}-\frac{\pi ^2}{24}-\frac{64}{27} \ln\frac{3}{c_{\star 0}}+\frac{1}{2} \ln^2\frac{3}{c_{\star 0}}\right) q_{\star 1} q_{\star 2}q_{\star 3}\nb\\
&&
+\left(\frac{611}{81}-4 \ln\frac{3}{c_{\star 0}}\right)q_{\star 1} q_{\star 2} \epsilon_{\star 1}+\left(\frac{19}{324}+\frac{\pi ^2}{24}+\frac{10}{27} \ln\frac{3}{c_{\star 0}}-\frac{1}{2} \ln^2\frac{3}{c_{\star 0}}\right) \epsilon_{\star 2} \epsilon_{\star 3} \epsilon_{\star 4}+\left(\frac{803}{81}-5 \ln\frac{3}{c_{\star 0}}\right) q_{\star 1} \epsilon_{\star 1} \epsilon_{\star 2}\nb\\
&&
+\left(-\frac{757}{81}+6 \ln\frac{3}{c_{\star 0}}\right) \epsilon_{\star 1}^2\epsilon_{\star 2}+\left(-\frac{55}{18}+\frac{\pi ^2}{12}+\frac{101}{27} \ln\frac{3}{c_{\star 0}}-\ln^2\frac{3}{c_{\star 0}}\right) \epsilon_{\star 1} \epsilon_{\star 2}^2
+\left(\frac{103}{81}-2 \ln\frac{3}{c_{\star 0}}\right) q_{\star 1} \epsilon_{\star 2} \epsilon_{\star 3}\nb\\
&&
+\left(-\frac{185}{54}+\frac{\pi^2}{12}+\frac{128}{27} \ln\frac{3}{c_{\star 0}}-\ln^2\frac{3}{c_{\star 0}}\right) \epsilon_{\star 1}\epsilon_{\star 2} \epsilon_{\star 3}+\mathcal{O}(\epsilon^4).
\eqn

Finally,  the running of the scalar spectral index can be determined up to the fourth-order in terms of  the slow-roll parameters, and is given by
\bqn
\alpha_s&\simeq&
q_{\star 1} q_{\star 2}-2 \epsilon_{\star 1} \epsilon_{\star 2}-\epsilon_{\star 2} \epsilon_{\star 3}-3 q_{\star 1}^2 q_{\star 2}+\left(\frac{64}{27}-\ln\frac{3}{c_{\star 0}}\right) q_{\star 1} q_{\star 2}^2+4 q_{\star 1} q_{\star 2} \epsilon_{\star 1}+5 q_{\star 1} \epsilon_{\star 1} \epsilon _{\star 2}-6 \epsilon_{\star 1}^2 \epsilon_{\star 2}+2 q_{\star 1} \epsilon_{\star 2} \epsilon_{\star 3}\nb\\
&&
+\left(2 \ln\frac{3}{c_{\star 0}}-\frac{101}{27}\right) \epsilon _{\star 1} \epsilon_{\star 2}^2+ \left(\frac{64}{27}-\ln\frac{3}{c_{\star 0}} \right)q_{\star 1}q_{\star 2} q_{\star 3}+q_{\star 1} q_{\star 2} \epsilon_{\star 2}+6 q_{\star 1}^3 q_{\star 2}+\left(2 \ln\frac{3}{c_{\star 0}}-\frac{128}{27}\right) \epsilon_{\star 1} \epsilon_{\star 2} \epsilon_{\star 3}\nb\\
&&
+\left(\ln\frac{3}{c_{\star 0}}-\frac{10}{27}\right)\epsilon_{\star 2} \epsilon_{\star 3}^2+\left(\ln\frac{3}{c_{\star 0}}-\frac{10}{27}\right) \epsilon _{\star 2} \epsilon_{\star 3} \epsilon_{\star 4}+\left(7 \ln\frac{3}{c_{\star 0}}-\frac{1076}{81}-\frac{64 \ln2}{315}\right)q_{\star 1}^2 q_{\star 2}^2-15 q_{\star 1}^2 q_{\star 2} \epsilon_{\star 1}\nb\\
&&
+\left(\frac{260}{81}-\frac{\pi ^2}{24}-\frac{64}{27} \ln\frac{3}{c_{\star 0}}+\frac{1}{2} \ln^2\frac{3}{c_{\star 0}}\right) q_{\star 1} q_{\star 2}^3
+\left(4 \ln\frac{3}{c_{\star 0}}-\frac{634}{81}-\frac{32 \ln2}{315}\right) q_{\star 1}^2 q_{\star 2}q_{\star 3}-3 q_{\star 1}^2 q_{\star 2} \epsilon_{\star 2}\nb\\
&&
+\left(\frac{260}{27}-\frac{\pi ^2}{8}-\frac{64}{9} \ln\frac{3}{c_{\star 0}}+\frac{3}{2} \ln^2\frac{3}{c_{\star 0}}\right)q_{\star 1} q_{\star 2}^2 q_{\star 3}
+\left(\frac{260}{81}-\frac{\pi ^2}{24}-\frac{64}{27} \ln\frac{3}{c_{\star 0}}+\frac{1}{2} \ln^2\frac{3}{c_{\star 0}}\right) q_{\star 1} q_{\star 2} q_{\star 3}^2\nb\\
&&
+\left(\frac{260}{81}-\frac{\pi ^2}{24}-\frac{64 \ln2}{27}+\ln^22-\ln2 \ln\frac{3}{c_{\star 0}}+\frac{1}{2} \ln^2\frac{3}{c_{\star 0}}\right) q_{\star 1} q_{\star 2}q_{\star 3} q_{\star 4}+\left(\frac{803}{81}-5 \ln\frac{3}{c_{\star 0}}\right) q_{\star 1} q_{\star 2}^2 \epsilon_{\star 1}\nb\\
&&
+\left(\frac{803}{81}-5 \ln\frac{3}{c_{\star 0}}\right) q_{\star 1} q_{\star 2} q_{\star 3} \epsilon_{\star 1}+9 q_{\star 1} q_{\star 2} \epsilon_{\star 1}^2
+\left(\frac{19}{324}+\frac{\pi ^2}{24}+\frac{10}{27}\ln\frac{3}{c_{\star 0}}-\frac{1}{2} \ln^2\frac{3}{c_{\star 0}}\right) \epsilon_{\star 2} \epsilon_{\star 3} \epsilon_{\star 4} \epsilon _{\star 5}\nb\\
&&
+\left(\frac{73}{81}-\ln\frac{3}{c_{\star 0}}\right) q_{\star 1} q_{\star 2}^2 \epsilon_{\star 2}+\left(\frac{73}{81}-\ln\frac{3}{c_{\star 0}}\right) q_{\star 1} q_{\star 2} q_{\star 3} \epsilon_{\star 2}-9 q_{\star 1}^2 \epsilon_{\star 1} \epsilon_{\star 2}+\left(\frac{1495}{81}-9 \ln\frac{3}{c_{\star 0}}\right) q_{\star 1} q_{\star 2} \epsilon_{\star 1} \epsilon_{\star 2}\nb\\
&&
+21 q_{\star 1} \epsilon_{\star 1}^2 \epsilon_{\star 2}-12 \epsilon_{\star 1}^3 \epsilon_{\star 2}+\left(\frac{1106}{81}-7 \ln\frac{3}{c_{\star 0}}\right) q_{\star 1} \epsilon_{\star 1} \epsilon_{\star 2}^2+\left(-\frac{1817}{81}+14 \ln\frac{3}{c_{\star 0}}\right) \epsilon_{\star 1}^2 \epsilon_{\star 2}^2\nb\\
&&
+\left(-\frac{55}{18}+\frac{\pi ^2}{12}+\frac{101}{27} \ln\frac{3}{c_{\star 0}}-\ln^2\frac{3}{c_{\star}}\right) \epsilon_{\star 1} \epsilon_{\star 2}^3-3 q_{\star 1}^2 \epsilon_{\star 2} \epsilon_{\star 3}+\left(\frac{176}{81}-3 \ln\frac{3}{c_{\star 0}}\right) q_{\star 1} q_{\star 2} \epsilon_{\star 2} \epsilon _{\star 3}\nb\\
&&
+\left(\frac{1349}{81}-7 \ln\frac{3}{c_{\star 0}}\right) q_{\star 1} \epsilon_{\star 1} \epsilon _{\star 2} \epsilon_{\star 3}+\left(-\frac{1141}{81}+8 \ln\frac{3}{c_{\star 0}}\right) \epsilon_{\star 1}^2 \epsilon_{\star 2} \epsilon_{\star 3}
+\left(\frac{133}{81}-3 \ln\frac{3}{c_{\star 0}}\right) q_{\star 1} \epsilon _{\star 2} \epsilon_{\star 3}^2
\nb\\
&&
+\left(-\frac{515}{54}+\frac{\pi ^2}{4}+\frac{110}{9} \ln\frac{3}{c_{\star 0}}-3 \ln^2\frac{3}{c_{\star 0}}\right) \epsilon_{\star 1} \epsilon _{\star 2}^2 \epsilon_{\star 3}+\left(\frac{19}{108}+\frac{\pi ^2}{8}+\frac{10}{9} \ln\frac{3}{c_{\star 0}}-\frac{3}{2} \ln^2\frac{3}{c_{\star 0}}\right) \epsilon_{\star 2} \epsilon_{\star 3}^2 \epsilon _{\star 4}\nb\\
&&
+\left(-\frac{205}{54}+\frac{\pi ^2}{12}+\frac{155}{27} \ln(\frac{3}{c_{\star 0}}-\ln^2\frac{3}{c_{\star 0}}\right)\epsilon_{\star 1} \epsilon_{\star 2} \epsilon _{\star 3}^2+\frac{76}{81} \epsilon_{\star 2}^2 \epsilon_{\star 3}^2+\left(\frac{19}{324}+\frac{\pi ^2}{24}+\frac{10}{27} \ln\frac{3}{c_{\star 0}}-\frac{1}{2} \ln^2\frac{3}{c_{\star 0}}\right) \epsilon_{\star 2} \epsilon_{\star 3}^3\nb\\
&&
+\left(\frac{133}{81}-3 \ln\frac{3}{c_{\star 0}}\right) q_{\star 1} \epsilon_{\star 2} \epsilon_{\star 3} \epsilon_{\star 4}+\left(-\frac{205}{54}+\frac{\pi ^2}{12}+\frac{155}{27} \ln\frac{3}{c_{\star 0}}-\ln^2\frac{3}{c_{\star 0}}\right) \epsilon_{\star 1} \epsilon_{\star 2}\epsilon_{\star 3} \epsilon_{\star 4}+\frac{38}{81} \epsilon_{\star 2}^2 \epsilon_{\star 3} \epsilon _{\star 4}\nb\\
&&
+\left(\frac{19}{324}+\frac{\pi ^2}{24}+\frac{10}{27} \ln\frac{3}{c_{\star 0}}-\frac{1}{2} \ln^2\frac{3}{c_{\star 0}}\right) \epsilon_{\star 2} \epsilon_{\star 3} \epsilon_{\star 4}^2
+\mathcal{O}(\epsilon^5).
\eqn

Now let us turn to consider the tensor perturbations. For the tensor spectrum, we get
\bqn
\Delta_t^2(k) &\simeq& \frac{181 H_\star^2}{36e^3 \pi^2} \Bigg\{1+\left(-\frac{496}{181}+\ln9\right) \epsilon _{\star 1}+\left(\frac{517}{543}-\frac{630 \ln3}{181}+2 \ln^23\right) \epsilon _{\star 1}^2\nb\\
&&~~~~~~~~~~~~+\left(-\frac{4636}{1629}+\frac{\pi ^2}{12}+\frac{496 \ln3}{181}-\ln^23\right) \epsilon _{\star 1} \epsilon _{\star 2}+\mathcal{O}(\epsilon^3)\Bigg\}.
\eqn

The tensor spectral index $n_t$, on the other hand, can be  expressed as
\bqn
n_t &\simeq&
-2 \epsilon _{\star 1}-2 \epsilon _{\star 1}^2+\left(2 \ln3-\frac{74}{27}\right) \epsilon _{\star 1} \epsilon _{\star 2}-2\epsilon _{\star 1}^3+\left(-\frac{676}{81}+6 \ln3\right) \epsilon _{\star 1}^2 \epsilon _{\star 2}\nb\\
&&+\left(-\frac{425}{162}+\frac{\pi ^2}{12}+\frac{74 \ln3}{27}-\ln^23\right) \epsilon _{\star 1} \epsilon_{\star 2}^2
+\left(-\frac{425}{162}+\frac{\pi ^2}{12}+\frac{74 \ln3}{27}-\ln^23\right) \epsilon _{\star 1} \epsilon _{\star 2} \epsilon _{\star 3}+\mathcal{O}(\epsilon^4),
\eqn
while  the running of the tensor spectral index is given by
\bqn
\alpha_t &\simeq &
-2 \epsilon _{\star 1} \epsilon _{\star 2}-6 \epsilon _{\star 1}^2 \epsilon _{\star 2}+\left(2 \ln3-\frac{74}{27}\right) \epsilon _{\star 1} \epsilon _{\star 2}^2+\left(2 \ln3-\frac{74}{27}\right) \epsilon _{\star 1} \epsilon _{\star 2} \epsilon _{\star 3}-12 \epsilon _{\star 1}^3 \epsilon _{\star 2}+\left(14 \ln3-\frac{1574}{81}\right) \epsilon _{\star 1}^2 \epsilon _{\star 2}^2\nb\\
&&
+\left(\frac{\pi ^2}{12}+\frac{74 \ln3}{27}-\ln^23-\frac{425}{162}\right) \epsilon _{\star 1} \epsilon _{\star 2}^3+\left(8 \ln3-\frac{898}{81}\right) \epsilon _{\star 1}^2 \epsilon _{\star 2} \epsilon _{\star 3}+\left(\frac{\pi ^2}{4}+\frac{74 \ln3}{9}-3 \ln^23-\frac{425}{54}\right) \epsilon _{\star 1} \epsilon _{\star 2}^2 \epsilon _{\star 3}\nb\\
&&
+\left(-\frac{425}{162}+\frac{\pi ^2}{12}+\frac{74 \ln3}{27}-\ln^23\right) \epsilon _{\star 1} \epsilon _{\star 2} \epsilon _{\star 3}^2+\left(-\frac{425}{162}+\frac{\pi ^2}{12}+\frac{74 \ln3}{27}-\ln^23\right) \epsilon _{\star 1} \epsilon _{\star 2} \epsilon _{\star 3} \epsilon _{\star 4}+\mathcal{O}(\epsilon^5).
\eqn

Finally, with both the scalar and tensor spectra given above, we can evaluate the tensor-to-scalar ratio at the time when tensor leaves horizon, i.e., $a(\eta_\star) H_\star=k$, and find that
\bqn
r&\simeq&
16 c_{\star0} \epsilon_{\star1} \Bigg\{1+\left(-\frac{429}{181}+\ln\frac{3}{c_{\star 0}}\right) q_{\star 1}+2 \epsilon _{\star 1} \ln c_{\star 0}+\left(\frac{67}{181}-\ln\frac{3}{c_{\star 0}}\right) \epsilon _{\star 2}\nb\\
&&+\left(\frac{285365}{65522}+\frac{64 \ln2}{1267}-\frac{610}{181} \ln\frac{3}{c_{\star 0}}+\frac{1}{2} \ln^2\frac{3}{c_{\star 0}}\right) q_{\star 1}^2+\left(-\frac{4865}{1629}+\frac{\pi ^2}{24}+\frac{429}{181} \ln\frac{3}{c_{\star 0}}-\frac{1}{2} \ln^2\frac{3}{c_{\star 0}}\right) q_{\star 1} q_{\star 2}\nb\\
&&
+\left(4\ln3+\left(\frac{1401}{181}+\ln9\right) \ln\frac{3}{c_{\star 0}}-2 \ln^2\frac{3}{c_{\star 0}}-\frac{508394}{98283}-\frac{1401 \ln3}{181}\right) q_{\star 1} \epsilon _{\star 1}+2 \left(\ln c_{\star 0}+\ln^2c_{\star 0}\right) \epsilon _{\star 1}^2\nb\\
&&
+\left(\frac{42500}{98283}+\frac{630 \ln3}{181}-\ln^23-\left(\frac{811}{181}+\ln9\right) \ln\frac{3}{c_{\star 0}}+3 \ln^2\frac{3}{c_{\star 0}}\right) \epsilon _{\star 1} \epsilon _{\star 2}+\left(\frac{1}{2} \ln^2\frac{3}{c_{\star 0}}-\frac{32615}{196566}-\frac{67}{181} \ln\frac{3}{c_{\star 0}}\right) \epsilon _{\star 2}^2\nb\\
&&
+\left(-\frac{174811}{98283}+\frac{677}{181} \ln\frac{3}{c_{\star 0}}-\ln^2\frac{3}{c_{\star 0}}\right) \epsilon _{\star 2} q_{\star 1}+\left(\frac{86}{1629}-\frac{\pi ^2}{24}-\frac{67}{181} \ln\frac{3}{c_{\star 0}}+\frac{1}{2} \ln^2\frac{3}{c_{\star 0}}\right) \epsilon _{\star 2} \epsilon _{\star 3}+\mathcal{O}(\epsilon^3)\Bigg\}.
\eqn

\section{Comparison with results obtained from other approximations}
\renewcommand{\theequation}{5.\arabic{equation}} \setcounter{equation}{0}

In this section, we compare the expressions we obtained in the above section with the ones obtained by other approximations, including the first-order uniform asymptotic approximation, the Green's function method, WKB approximation, and improved WKB approximation. To do so, we need to restrict ourselves to the special case $c_s(\eta)=1$.

Let us first consider the scalar spectrum. The scalar spectrum by the first-order uniform asymptotic approximation is given by
\bqn
\mathcal{P}_{\zeta,\text{1st-uniform}} &\simeq& \frac{18 H_{\star }^2}{8 e^3 \pi ^2 \epsilon _{\star 1}} \Bigg\{1+\left(-\frac{8}{3}+2 \ln3\right) \epsilon _{\star 1}+\left(-\frac{1}{3}+\ln3\right) \epsilon _{\star 2} +\left(\frac{7}{9}-\frac{10 \ln3}{3}+2 \ln^23\right) \epsilon _{\star 1}^2\nb\\
&&~~~~~~~~~~~~~~~+\left(-\frac{22}{9}+\frac{\pi ^2}{12}+\ln^23+\frac{\ln3}{3}\right) \epsilon _{\star 1} \epsilon _{\star 2}+\left(\frac{5}{18}-\frac{\ln3}{3}+\frac{\ln^23}{2}\right) \epsilon _{\star 2}^2\nb\\
&&~~~~~~~~~~~~~~~+\left(-\frac{1}{9}+\frac{\pi^2}{24}+\frac{\ln3}{3}-\frac{\ln^23}{2}\right) \epsilon _{\star 2} \epsilon _{\star 3}\Bigg\}.
\eqn
In Refs. \cite{Stewart2001PLB, Leach2002PRD}, the authors have found the scalar spectrum by using the Green function method, which is given by
\bqn\lb{scalarGreen}
\mathcal{P}_{\zeta, \text{Green}}&\simeq&\frac{H_\star^2}{8\pi^2 M_{\text{pl}}^2 \epsilon_{\star1}}
\Bigg\{1+2 (\alpha_\star-1) \epsilon_{\star1}+\alpha_\star \epsilon_{\star2}+\left(2\alpha^2_\star-2 \alpha_\star+\frac{\pi^2}{2}-5\right) \epsilon_{\star1}^2 \nb\\
&&~~~~~~~+\left(\alpha_\star^2+\alpha_\star+\frac{7\pi^2}{12}-7\right)\epsilon_{\star1} \epsilon_{\star2}+\left(\frac{\alpha_\star^2}{2}+\frac{\pi^2}{8}-1\right) \epsilon_{\star2}^2+\left(\frac{\pi^2}{24}-\frac{\alpha_\star^2}{2}\right)\epsilon_{\star2}\epsilon_{\star3}\Bigg\},
\eqn
where $\alpha_\star \equiv 2-\ln2-\gamma \simeq 0.729637$ with $\gamma$ being the Euler constant $\gamma \simeq 0.577216$. By using the WKB approximation, the scalar spectrum at second order is expressed as
\bqn
\mathcal{P}_{\zeta, \text{WKB}} &\simeq&\frac{H_\star^2}{8\pi^2 M_{\text{pl}}^2 \epsilon_{\star1}}  A_{\text{WKB}}
\Bigg\{1-2 (D_{\text{WKB}}-1) \epsilon_{\star1}-D_{\text{WKB}} \epsilon_{\star2}+\left(2D_{\text{WKB}}^2+2D_{\text{WKB}}-\frac{1}{9}\right) \epsilon_{\star1}^2 \nb\\
&&~~~+\left(D_{\text{WKB}}^2-D_{\text{WKB}}+\frac{\pi^2}{12}-\frac{20}{9}\right)\epsilon_{\star1} \epsilon_{\star2}+\left(\frac{D_{\text{WKB}}^2}{2}+\frac{\pi^2}{8}-1\right) \epsilon_{\star2}^2+\left(\frac{\pi^2}{24}-\frac{D_{\text{WKB}}^2}{2}\right)\epsilon_{\star2}\epsilon_{\star3}\Bigg\},\nb\\
\eqn
where $D_{\text{WKB}}=-\ln3+1/3 \simeq -0.765278955$ and $A_{\text{WKB}}=18 e^{-3} \simeq 0.896167$. In Ref.\  \cite{ImprovedWKB2005PRD}, the above results were further  improved by taking the next order in the adiabatic approximation into account. With such improvement, the scalar spectrum is given by
\bqn
\mathcal{P}_{\zeta, \text{WKB}*} &\simeq&\frac{H_\star^2}{8\pi^2 M_{\text{pl}}^2 \epsilon_{\star1}}  A_{\text{WKB}\star}
\Bigg\{1-2 (D_{\text{WKB}*}-1) \epsilon_{\star1}-D_{\text{WKB}*} \epsilon_{\star2}+\left(2D_{\text{WKB}*}^2+2D_{\text{WKB}*}-\frac{71}{1083}\right) \epsilon_{\star1}^2 \nb\\
&&~~~~~~~~~~~~~~~~~~~+\left(D_{\text{WKB}*}^2-D_{\text{WKB}*}+\frac{\pi^2}{12}+\frac{4}{57}b_S-\frac{2384}{1083}\right)\epsilon_{\star1} \epsilon_{\star2}+\left(\frac{D_{\text{WKB}*}^2}{2}+\frac{253}{1083}\right) \epsilon_{\star2}^2\nb\\
&&~~~~~~~~~~~~~~~~~~~+\left(\frac{\pi^2}{24}-\frac{D_{\text{WKB}*}^2}{2}+\frac{2}{57}b_S-\frac{49}{722}\right)\epsilon_{\star2}\epsilon_{\star3}\Bigg\},
\eqn
where $A_{\text{WKB}*}=361/(18e^3) \simeq 0.998507$, $D_{\text{WKB}*}=-\ln3+7/19 \simeq -0.730191$, and $b_S$ is an undetermined coefficient. For the overall amplitude in the above expressions, it is clear that the results obtained from the first-order uniform asymptotic approximation, the Green's function method, WKB approximation, and the improved WKB approximation, have a relative difference $\simeq 10.5\%,\;0.13\%,\;10.5\%$, or $0.28\%$, respectively, from ours that have an error bound $\lesssim 0.15\%$. It is worth emphasizing that the amplitudes in Eq.(\ref{scalarUniform}) and Eq.(\ref{scalarGreen}) are the closest ones among the four amplitudes.
In Table III, we also compare these numerical coefficients in the front of the Hubble flow parameters.

\begin{table}[htdp]
\caption{Comparison of the numerical coefficients of the scalar spectrum obtained by various methods}
\begin{center}
\begin{tabular}{|c|c|c|c|c|c|c|c|c|c|c|c|}
\hline
Method &  $\epsilon_{\star1}$ & $\epsilon_{\star2} $ &  $\epsilon_{\star1}^2$ &  $\epsilon_{\star1} \epsilon_{\star2} $ & $\epsilon_{\star2}^2$ & $ \epsilon_{\star2} \epsilon_{\star3}$\\
\hline
3rd-order uniform & -0.54310691 & 0.72844654 & -0.45788334 & 0.050725385 &0.4997524&0.16163455 \\
\hline
1st-order uniform & -0.46944209 & 0.76527896 & -0.47036526 & -0.048824354 & 0.51504816 & 0.062852021\\
\hline
Green function method  &-0.5407257 & -0.72963715 & -0.45973135 & 0.019276766 & 0.4998857 & 0.14504833  \\
\hline
WKB &-0.4694421 & 0.76527896 & -0.47036526 & -0.04882435 & 0.5150482 & 0.06285202  \\
\hline
improved WKB &-0.539617 & 0.7301912 & -0.459583 & -0.115455+0.070175 $b_S$ & 0.500200 & 0.07677686+0.035088 $b_S$  \\
\hline
\end{tabular}
\end{center}
\label{default}
\end{table}%


Now we turn to consider the tensor spectrum. The expressions from the first-order uniform asymptotic approximation, the Green's function method, WKB approximation, and  improved WKB approximation are given, respectively, by
\bqn
\mathcal{P}_{h,\text{1st-uniform}} &\simeq& \frac{36H_\star^2}{e^3 M^2_{\text{pl}}} \Bigg\{1+\left(2 \ln3-\frac{8}{3}\right) \epsilon _{\star 1}+\left(\frac{7}{9}-\frac{10 \ln3}{3}+2 \ln^23\right) \epsilon _{\star 1}^2\nb\\
&&~~~~~~~~~~~~~~~+\left(-\frac{26}{9}+\frac{\pi ^2}{12}+\frac{8 \ln3}{3}-\ln^23\right) \epsilon _{\star 1} \epsilon _{\star 2}\Bigg\},
\eqn
\bqn
\mathcal{P}_{h,\text{Green}} &\simeq& \frac{2 H_\star^2}{\pi^2 M_{\text{pl}}^2}
\Bigg\{1+2 (\alpha_\star-1) \epsilon_{\star1}+\left(2\alpha^2_\star-2 \alpha_\star+\frac{\pi^2}{2}-5\right) \epsilon_{\star1}^2 +\left(-\alpha_\star^2+2 \alpha_\star+\frac{\pi^2}{12}-2\right)\epsilon_{\star1} \epsilon_{\star2}\Bigg\},
\eqn
\bqn
\mathcal{P}_{h,\text{WKB}} &\simeq& \frac{2 H_\star^2}{\pi^2 M_{\text{pl}}^2} A_{\text{WKB}}
\Bigg\{1-2 (D_{\text{WKB}}+1) \epsilon_{\star1}+\left(2D_{\text{WKB}}^2+2 D_{\text{WKB}}-\frac{1}{9}\right) \epsilon_{\star1}^2 \nb\\
&&~~~~~~~~~~~~~~~~~~~~~-\left(D_{\text{WKB}}^2+2 D_{\text{WKB}}-\frac{\pi^2}{12}+\frac{19}{9}\right)\epsilon_{\star1} \epsilon_{\star2}\Bigg\},
\eqn
\bqn
\mathcal{P}_{h,\text{WKB}*} &\simeq& \frac{2 H_\star^2}{\pi^2 M_{\text{pl}}^2} A_{\text{WKB}*}
\Bigg\{1-2 (D_{\text{WKB}*}+1) \epsilon_{\star1}+\left(2D_{\text{WKB}*}^2+2 D_{\text{WKB}*}-\frac{1}{9}\right) \epsilon_{\star1}^2 \nb\\
&&~~~~~~~~~~~~~~~~~~~~~~-\left(D_{\text{WKB}*}^2+2 D_{\text{WKB}*}-\frac{\pi^2}{12}+\frac{19}{9}\right)\epsilon_{\star1} \epsilon_{\star2}\Bigg\}.
\eqn
In Table IV, we present the numerical coefficients of the tensor spectrum for each method.

\begin{table}[htdp]
\caption{Comparison of numerical coefficients of tensor spectrum}
\begin{center}
\begin{tabular}{|c|c|c|c|c|c|c|c|c|c|c|c|}
\hline
Methods &  $\epsilon_{\star1}$  &  $\epsilon_{\star1}^2$ &  $\epsilon_{\star1 }\epsilon_{\star2} $ \\
\hline
3rd-order uniform & -0.54310691  & -0.45788334 & -0.21983782  \\
\hline
1st-order uniform & -0.46944208 &  -0.47036526&  -0.34373805 \\
\hline
Green function method  &-0.5407257 & -0.45973135 & -0.25062903  \\
\hline
WKB method  &-0.4694421 & -0.47036526 & -0.34373805  \\
\hline
improved WKB method  & 0.539617&-0.459583  & -0.361440846  \\
\hline
\end{tabular}
\end{center}
\label{default}
\end{table}%

\end{widetext}

\section{Conclusions and Discussions}
\renewcommand{\theequation}{6.\arabic{equation}} \setcounter{equation}{0}

The uniform asymptotic approximation method provides a powerful, systematically improvable, and error-controlled approach to construct accurate analytical solutions of linear perturbations \cite{ZhuUniform2014PRD1,ZhuUniform2014-2}. In this paper, by applying the high-order uniform asymptotic approximations, we have obtained explicitly the analytical expressions of power spectra, spectral indices, and running of spectral indices for both  scalar and tensor perturbations in the $k$-inflation with the slow-roll approximation. These expressions are all written in terms of both the Hubble flow parameters defined in Eq.(\ref{Hubbleflow}) and sound speed flow parameters defined in Eq.(\ref{sound flow}). Comparing to the previous results obtained by the first-order uniform asymptotic approximation which in general have an error bound $\lesssim15\%$, the accuracy of the power spectra presented in this paper have been improved to $\lesssim 0.15\%$, which meets the accuracy of the current and forthcoming observations. In addition, the numerical coefficients in front of both the Hubble flow parameters and sound speed flow parameters are also highly improved by corrections from the high-order uniform asymptotic approximation.

Moreover, in Sec.\  V, we have made  detailed comparisons of the power spectra we obtained with the ones obtained by other approximate methods, including the first-order uniform asymptotic approximation, the Green's function method, WKB approximation, and  improved WKB approximation. It is shown that the results from the high-order uniform asymptotic approximation and the ones from the Green's function are the closest ones  among the results obtained by the five different methods.

\section*{Acknowledgements}

This work is supported in part by DOE, DE-FG02-10ER41692 (AW),
Ci\^encia Sem Fronteiras, No. 004/2013 - DRI/CAPES (AW),
NSFC No. 11375153 (AW), No. 11173021 (AW), No. 11047008 (TZ), No. 11105120 (TZ), No. 11205133 (TZ),
and URC/Baylor, No. 30330248 (QS).

\section*{Appendix A: Relations of slow-roll parameters}
\renewcommand{\theequation}{A.\arabic{equation}} \setcounter{equation}{0}

In this appendix, we present the relations between the Hubble flow parameters $\epsilon_i$ used in this paper and the slow-roll parameters $(\epsilon,\delta_i)$ used in Refs. \cite{Stewart2001PLB, ZhuUniform2014-2}. It can be shown that  the parameters $(\epsilon, \delta_i)$ in terms of the Hubble flow parameters $\epsilon_i$ are given by
\bqn
\epsilon &=&\epsilon_1,\nb\\
\delta_1 &=& \frac{\epsilon _2}{2}-\epsilon _1,\nb\\
\delta_2 &=& 2 \epsilon _1^2-\frac{5 \epsilon _2 \epsilon _1}{2}+\frac{\epsilon _2^2}{4}+\frac{\epsilon _2 \epsilon _3}{2}.
\eqn
On the other hand, the Hubble flow parameters can also be expressed in terms of $(\epsilon, \delta_i)$, and are given by
\bqn
\epsilon_1 &=& \epsilon,\nb\\
\epsilon_2 &=& 2 \delta _1+2 \epsilon,\nb\\
\epsilon_2 \epsilon_3 &=& 6 \delta _1 \epsilon -2 \delta _1^2+2 \delta _2+4 \epsilon ^2.
\eqn

\section*{Appendix B: Slow-roll expansions of $\nu_0,\;\nu_1,\;\nu_2,\;\nu_3$ and $c_1,\;c_2,\;c_3,\;c_4$}
\renewcommand{\theequation}{B.\arabic{equation}} \setcounter{equation}{0}

In this section we  present the results of the slow-roll expansions of the quantities $(\nu_0,\;\nu_1,\;\nu_2,\;\nu_3)$ and $(c_1,\;c_2,\;c_3,\;c_4)$. For the scalar perturbations, the expansions of $(\nu_0,\;\nu_1,\;\nu_2,\;\nu_3)$ are given, respectively, by
\bqn
\nu_0^{s}&\simeq& \frac{3}{2} + q_1+\epsilon _1+\frac{1}{2}\epsilon _2+q_1 \epsilon _1+\frac{1}{3}q_1 q_2+\epsilon _1^2+\frac{11}{6} \epsilon _2 \epsilon _1\nb\\
&&+\frac{1}{6}\epsilon _2 \epsilon _3+q_1 \epsilon _1^2+\frac{4}{9}q_1 q_2 \epsilon _1+\frac{10}{9}q_1 \epsilon _2 \epsilon _1-\frac{1}{9}q_1 q_2 \epsilon _2\nb\\
&&-\frac{1}{9} q_1 \epsilon _2 \epsilon _3-\frac{2}{9} q_1^2 q_2+\epsilon _1^3+\frac{77}{18} \epsilon _2 \epsilon _1^2+\frac{17}{9} \epsilon _2^2 \epsilon _1\nb\\
&&+\frac{14}{9} \epsilon _2 \epsilon _3 \epsilon _1-\frac{1}{18}\epsilon _2^2 \epsilon _3+\mathcal{O}(\epsilon^4),
\eqn
\bqn
\nu_1^{s} &\simeq & -q_1 q_2-\epsilon _1 \epsilon _2-\frac{\epsilon _2 \epsilon _3}{2}-2 q_1 q_2 \epsilon _1-q_1 \epsilon _1 \epsilon _2\nb\\
&&-\frac{1}{3} q_1 q_2^2-\frac{1}{3} q_1 q_3 q_2-\frac{11}{6} \epsilon _1 \epsilon _2^2-\frac{1}{6} \epsilon _2 \epsilon _3^2-3 \epsilon _1^2 \epsilon _2\nb\\
&&-\frac{7}{3} \epsilon _1 \epsilon _2 \epsilon _3-\frac{1}{6} \epsilon _2 \epsilon _3 \epsilon _4-3 q_1 q_2 \epsilon _1^2-3 q_1 \epsilon _2 \epsilon _1^2\nb\\
&&-\frac{7}{9} q_1 q_2^2 \epsilon _1-\frac{10}{9} q_1 \epsilon _2^2 \epsilon _1-\frac{7}{9} q_1 q_2 q_3 \epsilon _1-\frac{23}{9} q_1 q_2 \epsilon _2 \epsilon _1\nb\\
&&-\frac{10}{9} q_1 \epsilon _2 \epsilon _3 \epsilon _1+\frac{1}{9} q_1 \epsilon _2 \epsilon _3^2+\frac{1}{9} q_1 q_2^2 \epsilon _2+\frac{1}{9} q_1 q_2 q_3 \epsilon _2\nb\\
&&+\frac{2}{9} q_1 q_2 \epsilon _2 \epsilon _3+\frac{1}{9} q_1 \epsilon _2 \epsilon _3 \epsilon _4+\frac{4}{9} q_1^2 q_2^2+\frac{2}{9} q_1^2 q_2 q_3\nb\\
&&-6 \epsilon _2 \epsilon _1^3-\frac{205}{18} \epsilon _2^2 \epsilon _1^2-\frac{119}{18} \epsilon _2 \epsilon _3 \epsilon _1^2-\frac{17}{9} \epsilon _2^3 \epsilon _1\nb\\
&&-\frac{31}{18} \epsilon _2 \epsilon _3^2 \epsilon _1-\frac{35}{6} \epsilon _2^2 \epsilon _3 \epsilon _1-\frac{31}{18} \epsilon _2 \epsilon _3 \epsilon _4 \epsilon _1+\frac{1}{9} \epsilon _2^2 \epsilon _3^2\nb\\
&&+\frac{1}{18} \epsilon _2^2 \epsilon _3 \epsilon _4+\mathcal{O}(\epsilon^5),
\eqn
\bqn
\nu_2^{s}&\simeq& q_1 q_2^2+q_1 q_3 q_2+\epsilon _1 \epsilon _2^2+\frac{1}{2} \epsilon _2 \epsilon _3^2+\epsilon _1 \epsilon _2 \epsilon _3+\frac{1}{2} \epsilon _2 \epsilon _3 \epsilon _4\nb\\
&&+3 q_1 q_2^2 \epsilon _1+3 q_1 q_3 q_2 \epsilon _1+3 q_1 q_2 \epsilon _1 \epsilon _2+q_1 \epsilon _1 \epsilon _2^2+q_1 \epsilon _1 \epsilon _2 \epsilon _3\nb\\
&&+\frac{1}{3} q_1 q_2^3+q_1 q_3 q_2^2+\frac{1}{3} q_1 q_3^2 q_2+\frac{1}{3} q_1 q_3 q_4 q_2+\frac{11}{6} \epsilon _1 \epsilon _2^3\nb\\
&&+\frac{1}{6} \epsilon _2 \epsilon _3^3+7 \epsilon _1^2 \epsilon _2^2+\frac{17}{6} \epsilon _1 \epsilon _2 \epsilon _3^2+\frac{1}{6} \epsilon _2 \epsilon _3 \epsilon _4^2+6 \epsilon _1 \epsilon _2^2 \epsilon _3\nb\\
&&+4 \epsilon _1^2 \epsilon _2 \epsilon _3+\frac{1}{2} \epsilon _2 \epsilon _3^2 \epsilon _4+\frac{17}{6} \epsilon _1 \epsilon _2 \epsilon _3 \epsilon _4+\frac{1}{6} \epsilon _2 \epsilon _3 \epsilon _4 \epsilon _5\nb\\
&&+\mathcal{O}(\epsilon^5),
\eqn
\bqn
\nu_3^{s} &\simeq&-q_1 q_2^3-3 q_1 q_3 q_2^2-q_1 q_3^2 q_2-q_1 q_3 q_4 q_2-\epsilon _1 \epsilon _2^3\nb\\
&&-\frac{1}{2} \epsilon _2 \epsilon _3^3-\epsilon _1 \epsilon _2 \epsilon _3^2-\frac{1}{2} \epsilon _2 \epsilon _3 \epsilon _4^2-3 \epsilon _1 \epsilon _2^2 \epsilon _3-\frac{3}{2} \epsilon _2 \epsilon _3^2 \epsilon _4\nb\\
&&-\epsilon _1 \epsilon _2 \epsilon _3 \epsilon _4-\frac{1}{2} \epsilon _2 \epsilon _3 \epsilon _4 \epsilon _5+\mathcal{O}(\epsilon^5).
\eqn
For tensor perturbations, we find
\bqn
\nu_0^{t}&\simeq& \frac{3}{2}+\epsilon _1+\epsilon _1^2+\frac{4}{3} \epsilon _2 \epsilon _1+\epsilon _1^3+\frac{34}{9} \epsilon _2 \epsilon _1^2\nb\\
&&+\frac{4}{3} \epsilon _2^2 \epsilon _1+\frac{4}{3} \epsilon _2 \epsilon _3 \epsilon _1+\mathcal{O}(\epsilon^4),
\eqn
\bqn
\nu_1^{t} &\simeq& -\epsilon _1 \epsilon _2-3 \epsilon _2 \epsilon _1^2-\frac{4}{3} \epsilon _2^2 \epsilon _1-\frac{4}{3} \epsilon _2 \epsilon _3 \epsilon _1-6 \epsilon _2 \epsilon _1^3\nb\\
&&-\frac{89}{9} \epsilon _2^2 \epsilon _1^2-\frac{46}{9} \epsilon _2 \epsilon _3 \epsilon _1^2-\frac{4}{3} \epsilon _2^3 \epsilon _1-\frac{4}{3} \epsilon _2 \epsilon _3^2 \epsilon _1\nb\\
&&-4 \epsilon _2^2 \epsilon _3 \epsilon _1-\frac{4}{3} \epsilon _2 \epsilon _3 \epsilon _4 \epsilon _1+\mathcal{O}(\epsilon^5),
\eqn
\bqn
\nu^{t}_2 &\simeq& \epsilon _1 \epsilon _2^2+\epsilon _1 \epsilon _3 \epsilon _2+\frac{4}{3} \epsilon _1 \epsilon _2^3+7 \epsilon _1^2 \epsilon _2^2+4 \epsilon _1 \epsilon _3 \epsilon _2^2\nb\\
&&+\frac{4}{3} \epsilon _1 \epsilon _3^2 \epsilon _2+4 \epsilon _1^2 \epsilon _3 \epsilon _2+\frac{4}{3} \epsilon _1 \epsilon _3 \epsilon _4 \epsilon _2+\mathcal{O}(\epsilon^5),\nb\\
\eqn
\bqn
\nu^t_3 &\simeq& -\epsilon _1 \epsilon _2^3-3 \epsilon _1 \epsilon _3 \epsilon _2^2-\epsilon _1 \epsilon _3^2 \epsilon _2-\epsilon _1 \epsilon _3 \epsilon _4 \epsilon _2\nb\\
&&+\mathcal{O}(\epsilon^5).
\eqn

Now we turn to consider $c_1,\;c_2,\;c_3,\;c_4$, which are given by
\bqn
c_1 &\simeq& c_0\left(q_1 \epsilon _1^2+q_1 \epsilon _1+q_1 \epsilon _2 \epsilon _1+q_1+\mathcal{O}(\epsilon^4)\right),
\eqn
\bqn
c_2 &\simeq& c_0 \big(q_1^2-q_1 q_2+2 q_1^2 \epsilon _1-2 q_2 q_1 \epsilon _1-q_1 \epsilon _1 \epsilon _2+3 q_1^2 \epsilon _1^2\nb\\
&&~~~+2 q_1^2 \epsilon _1 \epsilon _2-3 q_2 q_1 \epsilon _1^2-q_1 \epsilon _1 \epsilon _2^2-3 q_1 \epsilon _1^2 \epsilon _2\nb\\
&&~~~-2 q_2 q_1 \epsilon _1 \epsilon _2-q_1 \epsilon _1 \epsilon _2 \epsilon _3+\mathcal{O}(\epsilon^5)\big),
\eqn
\bqn
c_3 &\simeq & c_0 \big(q_1^3-3 q_2 q_1^2+q_2^2 q_1+q_2 q_3 q_1+3 q_1^3 \epsilon _1-9 q_2 q_1^2 \epsilon _1\nb\\
&&\;\;\;-3 q_1^2 \epsilon _1 \epsilon _2+q_1 \epsilon _1 \epsilon _2^2+3 q_2^2 q_1 \epsilon _1+3 q_2 q_3 q_1 \epsilon _1\nb\\
&&\;\;\;+3 q_2 q_1 \epsilon _1 \epsilon _2+q_1 \epsilon _1 \epsilon _2 \epsilon _3+\mathcal{O}(\epsilon^5)\big),
\eqn
and
\bqn
c_4 &\simeq& c_0 \big(q_1^4-6 q_2 q_1^3+7 q_2^2 q_1^2+4 q_2 q_3 q_1^2-q_2^3 q_1-q_2 q_3^2 q_1\nb\\
&&\;\;\;-3 q_2^2 q_3 q_1-q_2 q_3 q_4 q_1+\mathcal{O}(\epsilon^5)\big).
\eqn

\end{document}